\documentclass[twocolumn,amsmath,amssymb,nofootinbib,a4paper]{revtex4} 
\usepackage[utf8]{inputenc}
\usepackage[T1]{fontenc}
\usepackage{newunicodechar}    
\usepackage{graphicx} 
\usepackage{amsmath}
\usepackage{amsfonts,amsbsy}
\usepackage{amssymb}
\usepackage{subdepth}
\usepackage{bbold}
\usepackage{nicefrac}
\usepackage{verbatim}
\usepackage{xcolor}
\definecolor{lcolor}{rgb}{0.5,0,0}
\definecolor{citcolor}{rgb}{0,0.3,0.0}
\usepackage[breaklinks,colorlinks,urlcolor=blue,citecolor=citcolor,linkcolor=lcolor]{hyperref}

\newcommand{\ktt}{k_\perp}

\newcommand{\nc}{{N_\mathrm{c}}}

\newcommand{\qs}{Q_\mathrm{s}}

\newcommand{\nr}[1]{(\ref{#1})} 

\newcommand{\fig}{Fig.~}
\newcommand{\figs}{Figs.~}
\newcommand{\eq}{Eq.~}
\newcommand{\se}{Sec.~}
\newcommand{\eqs}{Eqs.~}

\newcommand{\res}{Refs.~}

\begin{document}

\title{Plasmon mass scale in two-dimensional classical nonequilibrium gauge theory}

\author{T. Lappi}
\email{tuomas.v.v.lappi@jyu.fi}
\affiliation{
Department of Physics, %
 P.O. Box 35, 40014 University of Jyv\"askyl\"a, Finland
}
\affiliation{
Helsinki Institute of Physics, P.O. Box 64, 00014 University of Helsinki,
Finland
}

\author{J. Peuron}
\email{jarkko.t.peuron@student.jyu.fi}
\affiliation{
Department of Physics, %
 P.O. Box 35, 40014 University of Jyv\"askyl\"a, Finland
}

\begin{abstract}
We study the plasmon mass scale in classical gluodynamics in a two dimensional configuration that mimics the boost invariant initial color fields in a heavy ion collision. We numerically measure the plasmon mass scale using three different methods: a Hard Thermal Loop (HTL) expression involving the quasiparticle spectrum constructed from Coulomb gauge field correlators, an effective dispersion relation and the measurement of oscillations between electric and magnetic energies after introducing a spatially uniform perturbation to the electric field.
We find that the hard thermal loop expression and the uniform electric field measurement are in rough agreement. The effective dispersion relation agrees with other methods within a factor of 2.
We also study the dependence on time and occupation number, observing similar trends as in three spatial dimensions, where a power law dependence sets in after an occupation number dependent transient time. We observe a decrease of the plasmon mass squared as $t^{\nicefrac{-1}{3}}$ at late times. 
\end{abstract}

\maketitle

\section{Introduction}
The Color Glass Condensate (CGC) \cite{Iancu:2003xm,Gelis:2010nm} is an effective theory of Quantum ChromoDynamics (QCD) at high energy. One of the remarkable predictions of CGC is that two colliding CGC sheets create gluon states with nonperturbatively high occupation numbers ($\nicefrac{1}{g^2}$) \cite{Albacete:2014fwa}, which can be described using classical fields \cite{Lappi:2003bi}.

The question how this strongly interacting matter eventually isotropizes and thermalizes has been a longstanding question in the theory of ultrarelativistic heavy ion collisions. According to our current understanding the classical picture is valid for a  short time after the initial collision, until the occupation numbers fall below unity. The out of equilibrium matter admits a kinetic theory description when the occupation numbers are $\ll \nicefrac{1}{g^2}.$ Fortunately, the kinetic theory description also has an overlapping range of validity with classical simulations \cite{Mueller:2002gd,Jeon:2004dh,Berges:2004yj}, meaning that the two can be smoothly matched. In recent simulations the equilibration process has been studied by matching kinetic theory and relativistic hydrodynamics with promising results \cite{Kurkela:2015qoa}.

Calculations in the high collision energy limit, using the CGC formalism, predict that the initial color field configurations are boost invariant to leading order in the coupling \cite{Krasnitz:1998ns,Lappi:2009xa,Kovner:1995ts,Kovner:1995ja,Lappi:2006fp}. In practice the longitudinal structure of the colliding nuclei breaks this boost invariance at finite collision energies \cite{Gelfand:2016yho,Schenke:2016ksl,Ipp:2017lho}. In this paper we are, however, interested in the case where the boost invariace is only broken by small quantum fluctuations \cite{Fukushima:2006ax,Dusling:2010rm,Epelbaum:2011pc,Dusling:2012ig,Epelbaum:2013waa}. Due to instabilities, which are present in non abelian plasma \cite{Mrowczynski:1994xv,Mrowczynski:1996vh,Mrowczynski:2004kv,Romatschke:2003ms,
Kurkela:2011ub,Arnold:2003rq,Romatschke:2004jh,Arnold:2004ti,
Arnold:2004ih,Rebhan:2005re,Rebhan:2004ur,Kurkela:2011ti,Nara:2005fr,Dumitru:2005gp,
Bodeker:2007fw,Rebhan:2008uj,Attems:2012js}, even very small violations of boost invariance can grow rapidly, and become comparable to the classical background field. The plasma instability growth rate is determined by the plasmon mass scale \cite{Romatschke:2005pm,Romatschke:2006nk}, and therefore studying the plasmon mass can also shed light on the issue of isotropization and thermalization of plasma in ultrarelativistic heavy ion collisions.

The aim of this paper is to compare different methods to estimate the plasmon mass in classical Yang-Mills systems in the case where a 3-dimensional theory exists in a two-dimensional configuration. In practice we implement this by using a 3-dimensional calculation on a lattice with only one point in the $z$-direction. This configuration mimics the very anisotropic nearly boost-invariant field configuration predicted by the CGC for the initial stage of a heavy ion collision. At this stage we neglect the longitudinal expansion of the system in a heavy ion collision and work on a fixed size lattice. Instead of calculating in the asymptotically large time regime, where a clear separation between the hard and Debye scales has developed, we also want to address relatively early times where it is less obvious that such a scale separation exists. At this regime of earlier times  one can also study the dependence on the parameters of the initial condition, the occupation number and the hard scale, separately.

 To extract the plasmon mass we will systematically compare three methods, which we already compared in three dimensions \cite{Lappi:2016ato}. The first method will be to use a formula one can derive in Hard Thermal Loop (HTL) perturbation theory. Even though the HTL-like scale separation is not guaranteed by the weak coupling  in the classical theory, the plasmon mass scale nevertheless exists  \cite{Krasnitz:2000gz,Romatschke:2005pm,Romatschke:2006nk,Mace:2016svc,Mueller:2016ven}. The second method involves perturbing the system with a spatially uniform electric field (UE) \cite{Kurkela:2012hp}, and measuring the response to this zero momentum perturbation. The third method involves the effective dispersion relation (DR) \cite{Krasnitz:2000gz}, which we can extract from the Coulomb gauge correlators of the fields. 

The HTL formula relies on the quasiparticle spectrum, which is also typically extracted from Coulomb gauge correlators of the classical fields. However, we will notice that at high occupation numbers in two spatial dimensions the gauge fixing has a deforming effect on the observed quasiparticle spectrum. Thus we will argue for a need to use gauge invariant observables for measuring the typical momentum scale and occupation number.

We will first briefly introduce the numerical methods and initial conditions  in \se \ref{sec:method}. In \se \ref{sec:omegapl} we will introduce the three methods we use to extract the plasmon mass. Then we move on to dependencies on the lattice cutoffs, time and occupation number in \se \ref{sec:results}. Finally, we conclude in \se \ref{sec:conc}.

\begin{figure}[]
\centerline{\includegraphics[width=0.48\textwidth]{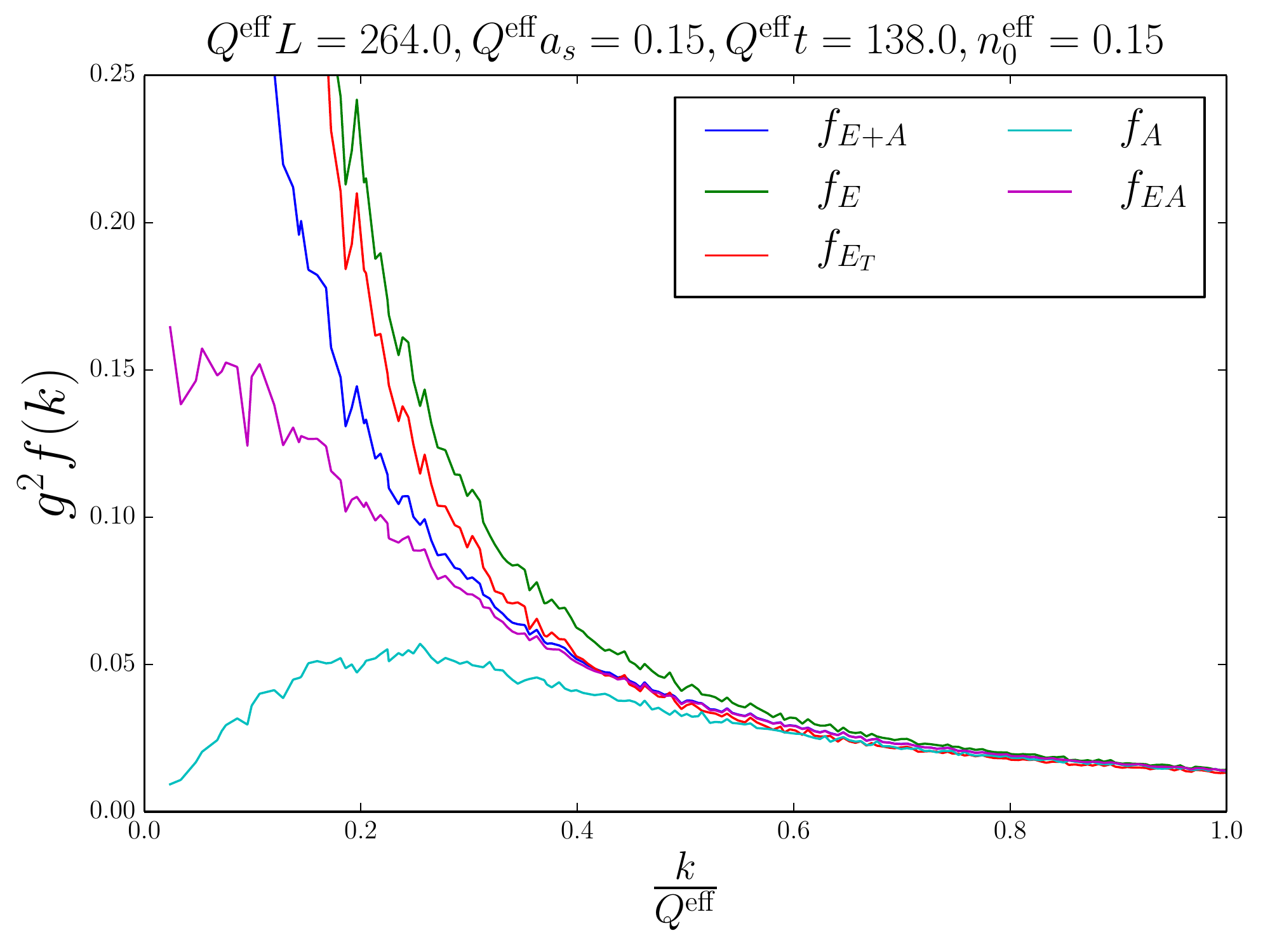}}
 \caption{Occupation number extracted using different methods.  Averaged over 20 configurations.}
 \label{fig:fmethods}
\end{figure}

\section{Numerical method and initial conditions}\label{sec:method}
\subsection{Equations of motion in the temporal gauge}
We have done all the numerical simulations with the SU(2) gauge group for numerical convenience. Several studies have compared SU(2) and SU(3) gauge groups, demonstrating qualitatively similar results \cite{Berges:2008zt,Ipp:2010uy,Berges:2017igc}.
We use the standard pure gauge Wilson action on a three-dimensional lattice 
\begin{align}
\mathcal{S} & =  -\beta_0 \sum_{x,i} \left( \frac{1}{N} \mathfrak{Re}\mathrm{Tr}\left(\underline{\Box}_x^{0,i}\right) - 	1 \right) \nonumber \\ & +  \beta_s \sum_{x,i<j} \left( \frac{1}{N} \mathfrak{Re}\mathrm{Tr}\left(\underline{\Box}_x^{i,j}\right) -1 \right), 
\label{eq:wilsonaction}
\end{align}
where $\beta_0 = \frac{2 N \gamma}{g^2},$ $\beta_s = \frac{2 N}{g^2 \gamma},$  $\gamma = \frac{a_s}{a_t},$ $N$ is the number of colors and $U_{x,i}$ are the link matrices defined as 
\begin{equation}
U_{x,i} = \exp{\left(i a_s g A_i\left(x\right) \right)}.
\label{eq:linkdef}
\end{equation}
 The plaquette variables appearing in (\ref{eq:wilsonaction}) are defined as $\underline{\Box}_x^{i,j} \equiv U_{x,i} U_{x+i,j} U^\dagger_{x+j,i} U^\dagger_{x,j }.$  The spatial lattice spacing is $a_s$ and the time step is denoted by $a_t.$ We use standard normalization for the SU(2) generators $\mathrm{Tr}\left(t^a t^b\right) = \dfrac{1}{2}\delta^{ab}.$ 

The relationships between the lattice variables and the actual fields are given by the following equations
\begin{equation}
E_i^a(x) = \dfrac{2}{a_s a_t g} \mathfrak{Im}\mathrm{Tr}(t^a\underline{\Box}_x^{i,0})
\end{equation}
\begin{equation}
B_i^a(x) = -\dfrac{ \varepsilon_{ijk}}{a_s^2 g} \mathfrak{Im}\mathrm{Tr} \left( t^a \underline{\Box}_x^{j,k} \right)
\end{equation}
\begin{equation}
 F_{\mu \nu}^a(x) = \dfrac{2}{ a_\mu a_\nu g} \mathfrak{Im}\mathrm{Tr} \left( t^a \underline{\Box}_x^{\mu,\nu} \right)
 \end{equation}
\begin{equation}
 A_\mu^a(x) = \dfrac{2}{a_\mu g } \mathfrak{Im} \mathrm{Tr} \left( t^a U_{x,\mu} \right).
\label{eq:Aextraction} 
 \end{equation}
 We refer to the lattice spacing in the $\mu$ direction with  $a_\mu.$
The equations of motion of the electric field are obtained by varying the action (\ref{eq:wilsonaction}) with respect to the spatial links
\begin{align}
\label{eq:Eevolution}
E_j (t , x)  & =  E_j(t-a_t , x) + \dfrac{a_t}{2 i a_s^3 g} \sum_k \Bigg(  \underline{\Box}_x^{j,k}   - \underline{\Box}_x^{k,j} \nonumber \\ &- \dfrac {\mathbb{1}}{N} \mathrm{Tr} \left(  \underline{\Box}_x^{j,k}   - \underline{\Box}_x^{k,j} \right) 
+ \overline{\Box}_x^{j,k} - \left(\overline{\Box}_x^{j,k}\right)^\dagger  \nonumber \\ & - \dfrac{\mathbb{1}}{N} \mathrm{Tr} \left( \overline{\Box}_x^{j,k} - \left(\overline{\Box}_x^{j,k}\right)^\dagger \right) \Bigg),  
\end{align}
where $\overline{\Box}_x^{j,k} = U_{x,j} U_{x+j-k,k}^\dagger U_{x-k,j}^\dagger U_{x-k,k}.$ 
When the electric field on the next time step is known, we can easily construct the temporal plaquette (when using the SU(2) symmetry group) by decomposing the temporal plaquette into two parts
\begin{equation}
\underline{\Box}_x^{i,0} = \sqrt{1 - \left(\dfrac{a_s a_t g}{2} E_a \right)^2}\mathbb{1} + ia_s a_t g E^a t^a.
\end{equation}
In the temporal gauge the temporal plaquette simplifies to a product of link matrices at two different time steps, making it easy to solve for the link at the next time step.
Color charge conservation is encoded in Gauss's law
\begin{equation}
\sum_j \left( E_j(x) - U_{x-j, j }^\dagger E_j(x-j) U_{x-j, j } \right) = 0,
\label{eq:gauss}
\end{equation}
which is preserved by the discretization algorithm.

\subsection{Quasiparticle distribution}\label{sec:partdist}
We extract the quasiparticle spectrum by eliminating the residual gauge freedom with the Coulomb gauge condition. The gauge fixing is done by a Fourier accelerated algorithm~\cite{Davies:1987vs}. However, even with gauge fixing, there is no unique way to determine a quasiparticle distribution from a given classical field configuration (see also the discussion in Refs.~\cite{Kurkela:2012hp} and \cite{Li:2017iat}). If our system can be described by weakly interacting quasiparticles, the energy density of the system should be obtained as 
\begin{equation} \label{eq:edensity2d}
\epsilon =  2\left(\nc^2-1\right)\int \dfrac{\mathrm{d}^3 k}{\left(2 \pi \right)^3} \omega\left(k\right) f\left(k\right).
\end{equation}
Here the factor $2\left(\nc^2-1\right)$ accounts for the number of color and transverse polarization states in the system. The number of physical polarization states the plasmons have is 3. However, the longitudinal mode is only present for modes close to the Debye scale, and is not expected to contribute significantly to the total energy density in (\ref{eq:edensity2d}). Thus a factor 3 would lead to a significant underestimation of the occupation number of hard modes.
The total energy of the system is given by the Hamiltonian
\begin{equation}
\mathcal{H} = \int \mathrm{d}^3x \mathrm{Tr}\left( E_i E^i + B_i B^i\right).
\end{equation}
We now keep only the terms which are quadratic in the gluon field and equate the energy given by the Hamiltonian with the one given by the quasiparticle spectrum. Solving for the quasiparticle spectrum gives 
\begin{equation} \label{eq:discF}
f_{A+E}\left(k\right) = \dfrac{1}{4\left(\nc^2-1 \right)} \dfrac{1}{V} \left(\dfrac{\left| E_C\left(k\right)\right|^2}{\omega\left(k\right)} + \dfrac{k^2}{\omega\left(k\right)} \left|A_C\left(k\right)\right|^2\right).
\end{equation}
Here $\left| E_C\left(k\right)\right|$ is the Coulomb gauge electric field and $\left| A_C\left(k\right)\right|$ is the gauge field in Coulomb gauge. This procedure also removes the magnetic part of the longitudinal polarization state.
The energy of a mode with momentum $k$ (the dispersion relation) is given by $\omega\left(k\right)$. When extracting the quasiparticle spectrum we will assume a massless dispersion relation $\omega(k)=k$. It is not immediately obvious what would be the correct procedure to self-consistently include a plasmon mass in the dispersion relation used here. In any case the effect of such a correction would be of the same order as the higher order terms in the gauge potential that we are already neglecting in \eq (\ref{eq:discF}).

Alternatively we can also extract the quasiparticle spectrum using only the gauge fields, or only electric fields. These two should be equivalent above the Debye scale after the system has been evolved in time for a few $Qt$. The alternative definitions for the occupation number are 
\begin{equation} \label{eq:discFE}
f_E\left(k\right) =  \dfrac{1}{2\left(\nc^2-1 \right)} \dfrac{1}{V} \left(\dfrac{\left| E_C\left(k\right)\right|^2}{\omega\left(k\right)} \right)
\end{equation}
for the electric estimator and
\begin{equation} \label{eq:discFA}
f_A\left(k\right) =  \dfrac{1}{2\left(\nc^2-1 \right)} \dfrac{1}{V} \left( \dfrac{k^2}{\omega\left(k\right)} \left|A_C\left(k\right)\right|^2\right)
\end{equation}
for the magnetic estimator. We can also use a combination of electric and magnetic fields
\begin{equation} \label{eq:discFEA}
f_{EA}\left(k\right) =  \dfrac{1}{2\left(\nc^2-1 \right)} \dfrac{1}{V} \left( \sqrt{\left|A_C\left(k\right)\right|^2\left| E_C\left(k\right)\right|^2}\right).
\end{equation}
We compare these different definitions for the occupation number in \fig \ref{fig:fmethods}. We observe that the electric occupation number diverges in the infrared, whereas the magnetic estimator is IR finite. The combination of electric and magnetic fields ($f_{EA}$) also behaves reasonably well in the IR.  This behavior is to be expected, since below the Debye scale we would not expect the dispersion relation to be massless. We will return to this question later in the context of using these distributions.  Figure~\ref{fig:fmethods} also demonstrates the transverse electric occupation number. We clearly see that it is very close to the total electric occupation number, especially at higher momenta, indicating that the quasiparticle spectrum is dominated by transverse plasmons.

\subsection{Initial conditions}
The initial gauge fields are sampled from the distribution 
\begin{equation}
\left< A_i^a\left(k\right) A_j^b\left(p\right)\right> = \dfrac{V n_0}{g^2 Q} \exp{\left(\dfrac{-k^2}{ Q^2}\right)} \delta_{ij} \delta^{ab} \dfrac{\delta^{\left(3\right)}\left(k+p\right) \left(2 \pi \right)^3}{V}.
\label{eq:IC}
\end{equation}
Since our system is two-dimensional, the $z$-components of the momenta are in fact always zero, and the exponential is $\exp(-k^2/ Q^2) = \exp(-\ktt^2/Q^2)$.
Here $Q$ is the dominant momentum scale, $V$ is the lattice volume and $n_0$ is a parameter describing the typical occupation number of the system. This initial condition is a momentum distribution clearly peaked around $Q$ and it behaves very well in the ultraviolet and infrared regions. It also  trivially satisfies Gauss's law, since it contains only magnetic energy. In the context of the early stages of heavy-ion collisions we should consider $Q$ as analogous to the saturation scale $\qs$ \cite{Gyulassy:2004zy}.
For the classical approximation to be valid, the occupation number $f\sim n_0/g^2$ should be greater than of the order of 1, i.e. $n_0>> g^2$.

Our choice of initial fields, using \nr{eq:discF} as the quasiparticle distribution, results in the following form~\footnote{Since we are starting from a configuration with purely magnetic energy, $f_A=2f_{A+E}$ at the initial time. After a decoherence time $\sim 1/Q$ approximately half of the energy moves to the electric sector. In a noninteracting theory this keeps $f_{A+E}$ the same and reduces $f_A$ by a factor 2.  Thus we use here $f_{A+E}$ from \nr{eq:discF}, since it gives  a better estimate of the total quasiparticle distribution inserted into the system by the initial condition.} \begin{equation}
f\left(k, t=0 \right) = \frac{n_0}{2 g^2} \dfrac{\ktt}{Q} \exp{\left( \dfrac{-\ktt^2}{2 Q^2}\right)}
 \frac{(2\pi)\delta(k_z)}{a_s},
\label{eq:initdist}
\end{equation}
where $a_s$ is the length of the system in the $z$-direction.

We are using the same initial condition as in Refs.~\cite{Berges:2007re,Berges:2012ev}. This distribution is strongly cut off in the UV and is in this regard similar to initial conditions used in Refs.~ \cite{Berges:2013eia,Berges:2014bba}. For more realistic initial conditions we refer the reader to, for example Refs.~\cite{Lappi:2011ju,Berges:2012cj}.
The precise functional form of the initial condition does not very strongly affect the late time behaviour of the system (unless one changes the occupation number by very large margin), because overoccupied classical Yang-Mills systems eventually evolve into a well known scaling solution in a time of the order of a  few $Qt$ \cite{Kurkela:2012hp, Berges:2013fga,Berges:2013eia}. Similar findings have also been made in scalar field theories \cite{Berges:2013lsa,Berges:2014bba,Berges:2015ixa}.

\subsection{Measured observables}
\begin{figure}[]
\centerline{\includegraphics[width=0.48\textwidth]{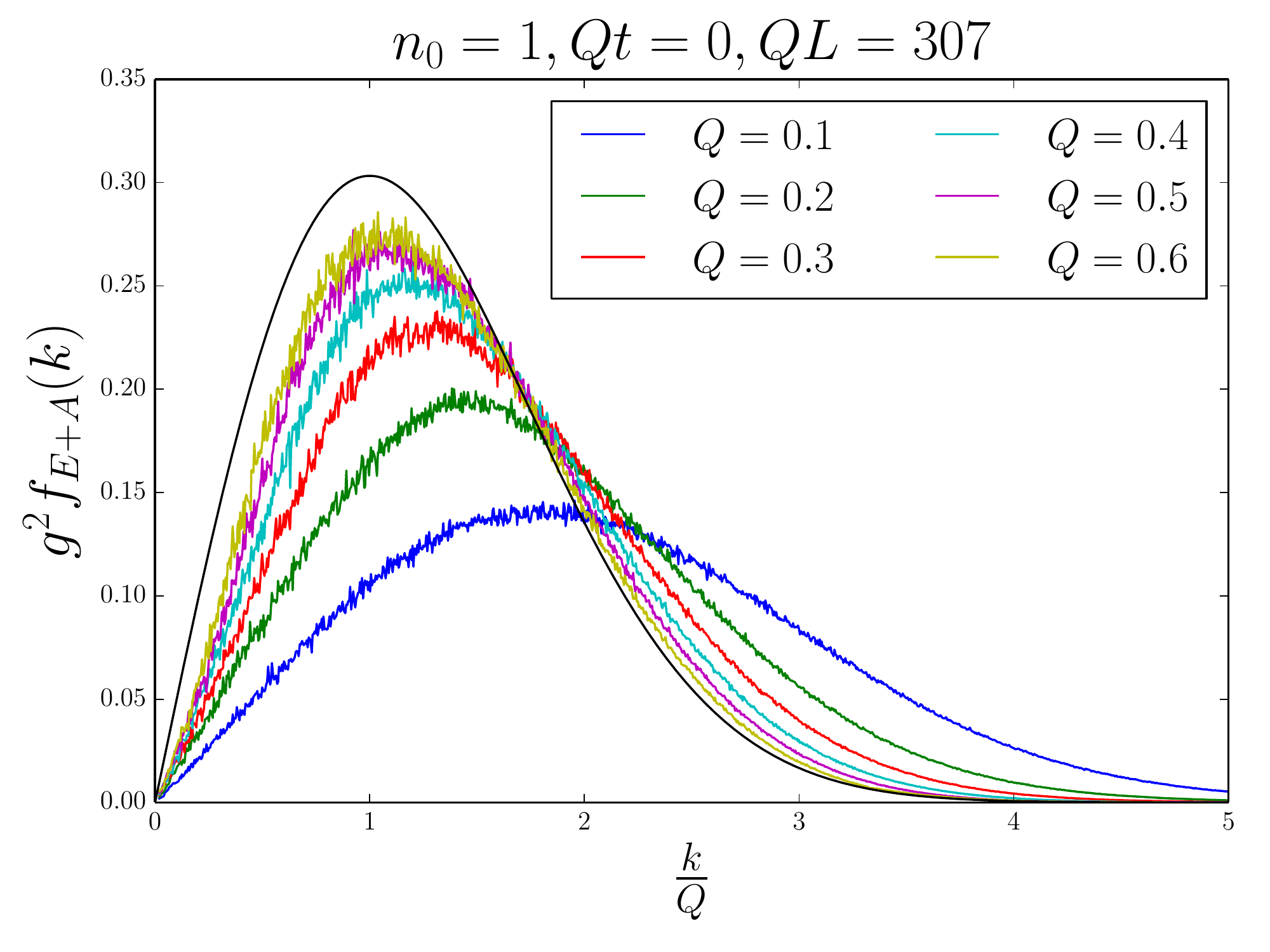}}
 \caption{The effect of the initial momentum scale on the observed quasiparticle spectrum. The configurations which are closest to the continuum deviate the most from the analytical form~\nr{eq:initdist} used as an input, shown as the smooth black line. Averaged over 10 runs.}
\label{fig:fvsQ}
\end{figure}

\begin{figure}[]
\centerline{\includegraphics[width=0.48\textwidth]{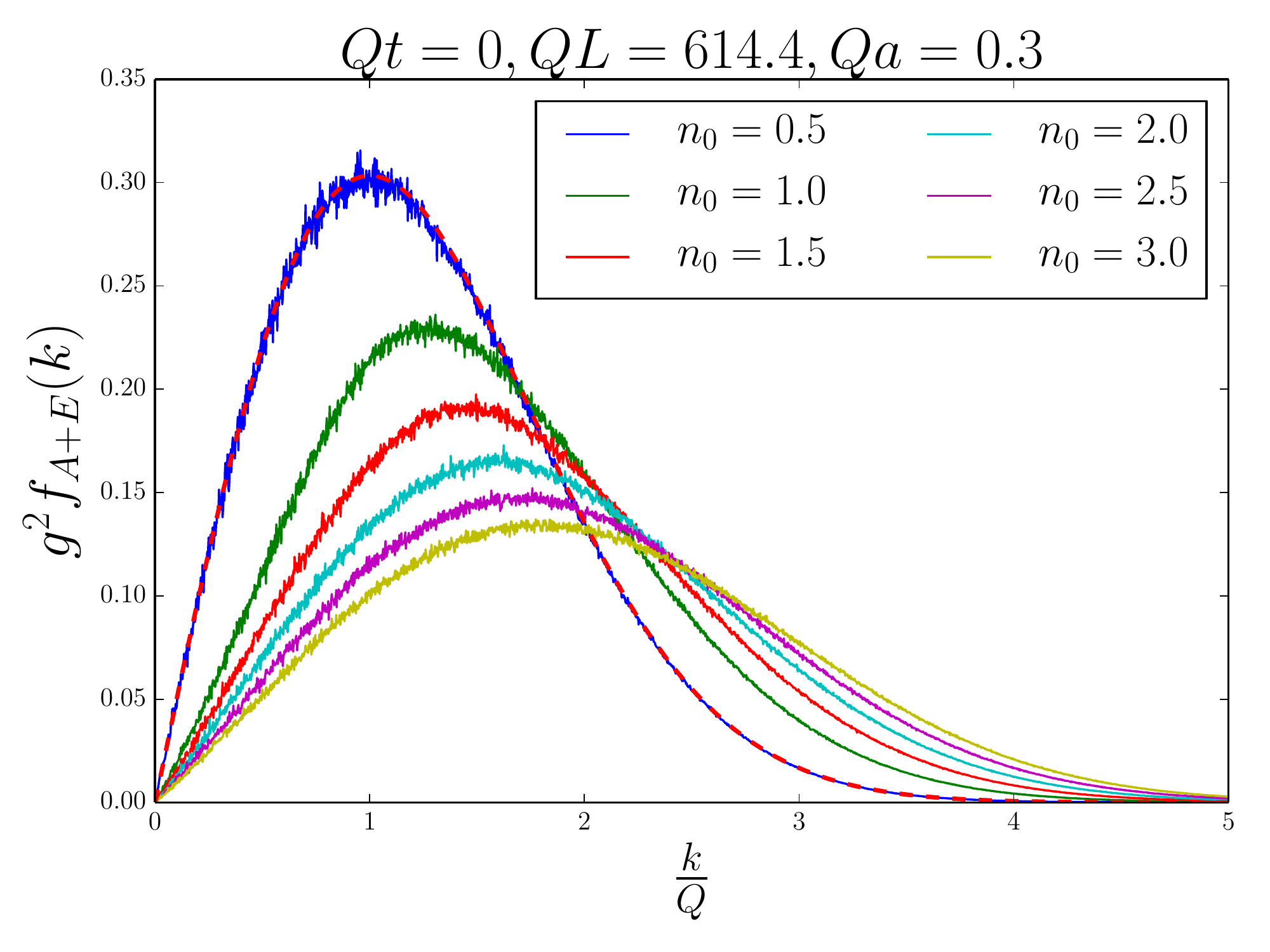}}
 \caption{The effect of the initial occupation number on the observed quasiparticle spectrum. For the larger occupation numbers the gauge fixing has the most dramatic effect on the quasiparticle spectrum. The analytical form~\nr{eq:initdist} used as an input, shown as the dashed red line, is barely visible under the $n_0=0.5$ curve.  Averaged over 10 runs.}
\label{fig:fvsn0}
\end{figure}

Our initial condition~\nr{eq:IC} is constructed with the gauge fields in momentum space. These are then Fourier transformed back to coordinate space and exponentiated to form the link matrices that are the actual variables in the calculation. In order to measure the quasiparticle spectrum~\nr{eq:discF} one then fixes the Coulomb gauge and calculates the antihermitian traceless part of the link matrix to get the gauge potential $A_i$ appearing in the expression~\nr{eq:discF}. This process is very nonlinear, and in the high density regime it is not obvious that one recovers the same quasiparticle distribution that one put in as an initial condition. In order to control the limits in this process it is useful to compare the measured quasiparticle distribution at $t=0$ to the one used in the initial condition. This comparison is shown in \figs \ref{fig:fvsQ} and \ref{fig:fvsn0}. Figure~\ref{fig:fvsQ} shows the dependence of the gauge fixed quasiparticle spectrum on the initial momentum scale. We observe that the gauge fixed spectrum deviates the most from the analytical initial condition when $Qa_s$ is small. Figure~\ref{fig:fvsn0} shows the gauge fixed spectrum for different occupation numbers. Here we find a similar effect, at higher occupancy the  gauge fixing significantly drives the spectrum away from the analytical initial condition by redistributing the energy to higher momentum modes and decreasing the occupation number correspondingly. This effect seems to be more dramatic in two dimensions than in three dimensions (In three dimensions we did not observe this effect with $n_0 = 1$, but we presume that this effect can be observed also in three dimensions, provided that one goes to high enough occupation numbers).

In order to reduce the gauge fixing effects, we project out the longitudinal components of the initial gauge field using the standard transverse projection operator
\begin{equation}
P^{ij}_{T} =  \delta_{ij} - \dfrac{p_{i}^* p_{j,B}}{\left|p\right|^2}.
\end{equation}
Here $p_i$ and $p_j$ correspond to the (complex) eigenvalues of the discretized derivative operator. However, the antihermitian parts of the links constructed by exponentiation from the transverse fields do not necessarily satisfy the Coulomb gauge condition to the desired accuracy. This problem is especially severe when the occupation numbers are high. The effect is also visible in the \fig \ref{fig:fvsn0}. The curve with $n_0=0.5$ overlaps with the analytical initial condition, because here the transverse projection done on the initial gauge field also keeps the lattice Coulomb gauge violation so small that no additional gauge fixing is needed. However, when one goes to higher occupation numbers, additional gauge fixing becomes necessary and we start to observe deviations from the analytical distribution function.

\begin{figure}[]
\centerline{\includegraphics[width=0.48\textwidth]
{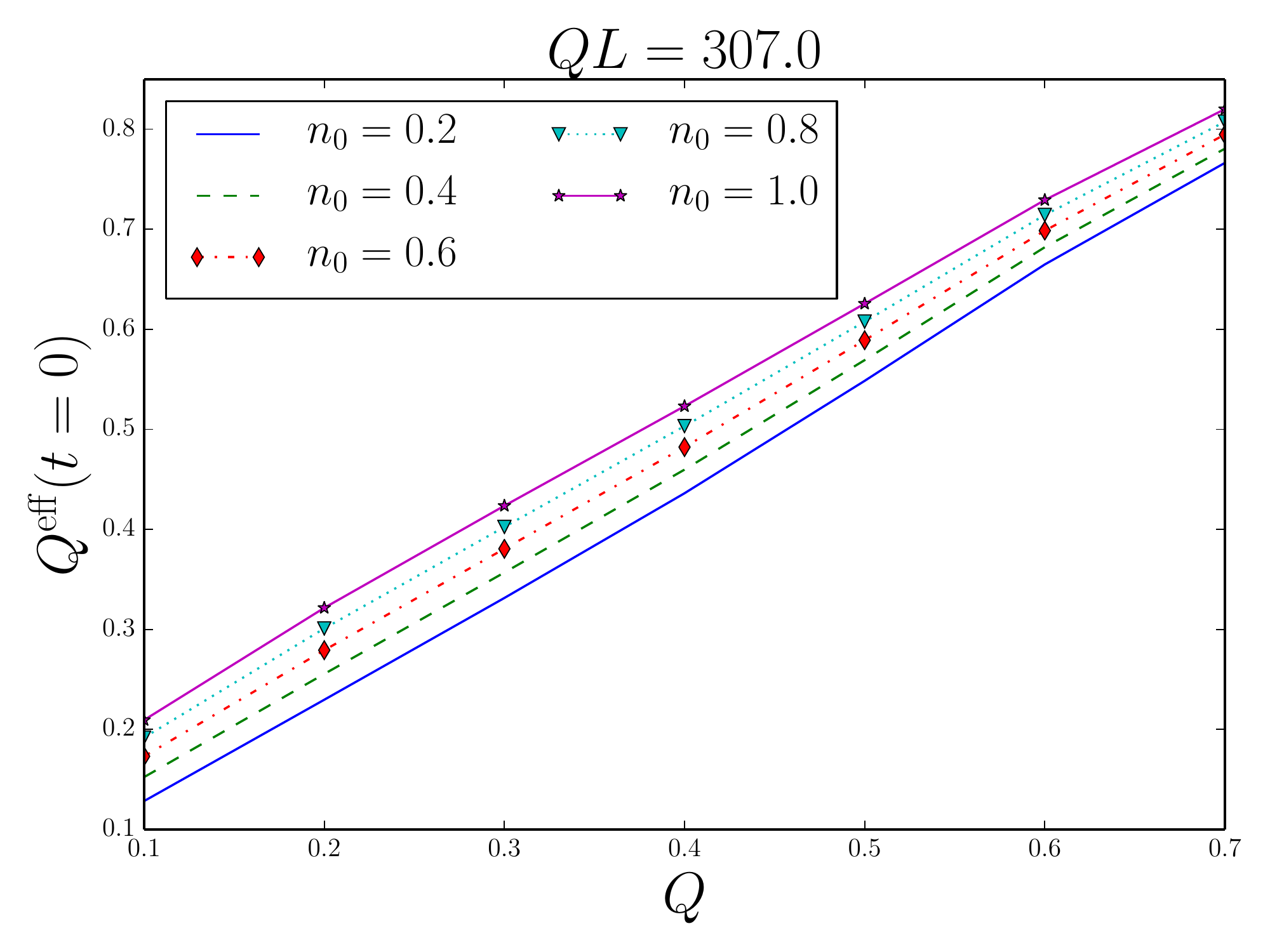}}
\caption{Effective momentum scale as a function of the scale set by the initial condition for different $n_0.$ We observe a linear dependence, but increasing $n_0$ also increases the observed $Q^{\mathrm{eff}}.$ }
\label{fig:QeffvsQ}
\end{figure}

\begin{figure}[]
\centerline{\includegraphics[width=0.48\textwidth]
{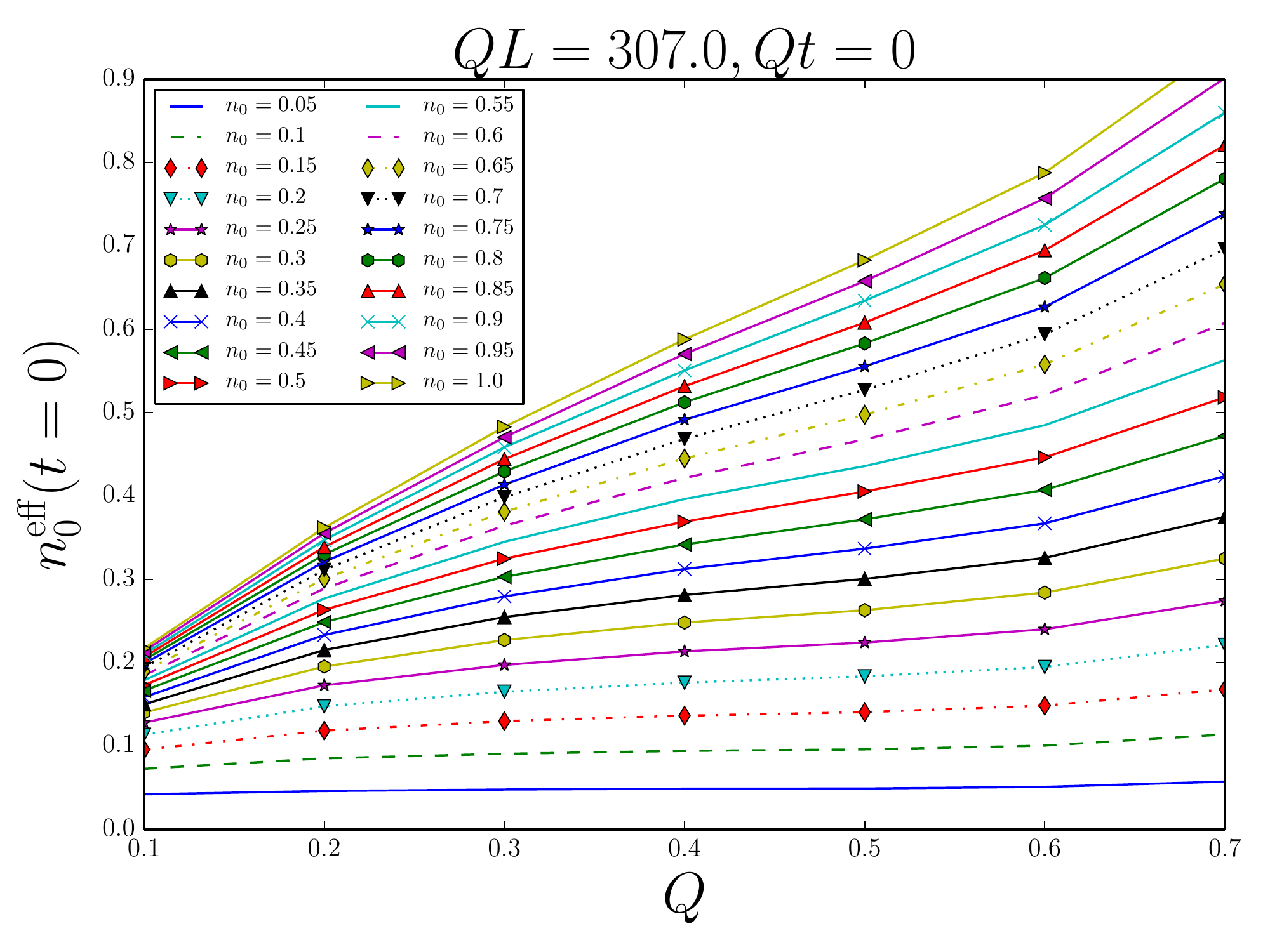}}
\caption{Effective occupation number as a function of the momentum scale set by the initial condition for different $n_0$. The lowest curves with $n_0^{\mathrm{eff}}\approx n_0$ independently of $Q$ correspond to the lowest $n_0$ values. All the curves are ordered in such a way that when $n_0$ becomes larger, also the $n_0^{\mathrm{eff}}$ becomes larger, averaged over five configurations. }
\label{fig:n0effvsQ}
\end{figure}

Due to this fact we want to measure the momentum scale and occupation number in a gauge invariant manner. We estimate the typical momentum of the chromomagnetic field squared as done in \cite{Bodeker:2007fw,Kurkela:2012hp} 
\begin{equation}
p_{\mathrm{eff}}^2(t) = \dfrac{\left< \mathrm{Tr}\left(\boldsymbol{D} \times \boldsymbol{B} \right)^2 \right>}{\left< \mathrm{Tr} \left(\boldsymbol{B}^2\right) \right>}.
\end{equation}
For our initial condition we can estimate this perturbatively 
\begin{equation}
p^2_{\mathrm{eff}}(t=0) \approx \dfrac{\int \mathrm{d}\ktt \ktt^4 f(\ktt)}{\int \mathrm{d}\ktt \ktt^2 f(\ktt)} = 4Q^2.
\end{equation}
We define the effective momentum scale in such a way that it matches the initial $Q$ in the dilute limit
\begin{equation}
Q_{\mathrm{eff}} = \dfrac{p_{\mathrm{eff}}}{2}.
\end{equation}
In order to estimate the occupation number, we first compute the initial energy density in terms of $Q$ and $n_0$ using \eq \nr{eq:edensity2d} 
\begin{equation}
\epsilon \approx  n_0 Q^3 \dfrac{(\nc^2-1)}{\pi}\frac{1}{a_s g^2}.
\end{equation}
This means that we can use the gauge invariant momentum scale $Q_{\mathrm{eff}}$ and the 2-dimensional energy density $\epsilon^{2d}\equiv a_s\epsilon$ to define a gauge invariant measure of the typical occupation number of gluons as %
\begin{equation}
n_0^{\mathrm{eff}} \approx \dfrac{\pi g^2}{(\nc^2-1)} \dfrac{\epsilon^{2d}}{Q^3} \approx \dfrac{\pi g^2}{(\nc^2-1)} \dfrac{8 \epsilon^{2d}}{ p_{\mathrm{eff}}^3}.
\end{equation}

The normalization of $Q_{\mathrm{eff}}$ and $n_0^{\mathrm{eff}}$ has now been chosen in such a way that in the dilute limit and at $t=0$ they agree with the input parameters $Q$ and $n_0$. Away from the dilute limit we want to perform simulations by varying the lattice parameters in a way that maintains as much as possible  fixed values of the gauge invariant scales $Q_{\mathrm{eff}}$ and $n_0^{\mathrm{eff}}$. In order to do  this, we must map the relation between the input parameters and the gauge invariant scales. To this end we have performed a series of measurements at the initialization time. Figure~\ref{fig:QeffvsQ} shows the dependence of the gauge invariant scale $Q^{\mathrm{eff}}$ on $Q$ for various occupation numbers $n_0$. The gauge invariant scale is linear in the initial scale, but an increase in the initial occupation number $n_0$ also results in an increased gauge invariant scale. This reinforces our interpretation of the phenomenon seen in the gauge-fixed spectra in \fig~\ref{fig:fvsn0}, where we observed that, for a fixed input parameter $Q$, the resulting effective  momentum scale increases when one increases the occupation number. It seems that the system resists attempts to increase the gluon density by increasing the amplitude $n_0$, instead transferring the additional energy into modes with higher momenta. 
Figure \ref{fig:n0effvsQ} shows the connection between the gauge invariant occupation number and the occupation number $n_0$ as a function of $Q$. The main conclusion of this figure is that at low $Qa_s$ it is impossible to go to high occupation numbers, but instead $n_0^{\mathrm{eff}}$ saturates to certain value. At higher $Q$ we can go to higher effective occupation numbers, but we are also further away from the continuum limit. 
It is also useful to  compare \figs \ref{fig:n0effvsQ} and \ref{fig:fvsQ}. When keeping $n_0$ fixed in \fig \ref{fig:n0effvsQ} and going to higher $Q$ we observe that  the effective occupation number also increases. Thus the behavior of the gauge fixed spectrum in \fig \ref{fig:fvsQ} is in line with the behavior we find in gauge invariant occupation number in \fig \ref{fig:n0effvsQ}.
To summarize, with the  help of \figs \ref{fig:QeffvsQ} and \ref{fig:n0effvsQ} we can establish a link between the initial simulation parameters, and the initial measured parameters. However, the reader should bear in mind that for example, taking the continuum limit ($Q^{\mathrm{eff}} a_s \to 0$ for fixed $n_0^{\mathrm{eff}}$) in the high occupation number regime is actually impossible due to the saturation in $n_0^{\mathrm{eff}}$.

\begin{figure}[]
\centerline{\includegraphics[width=0.48\textwidth]
{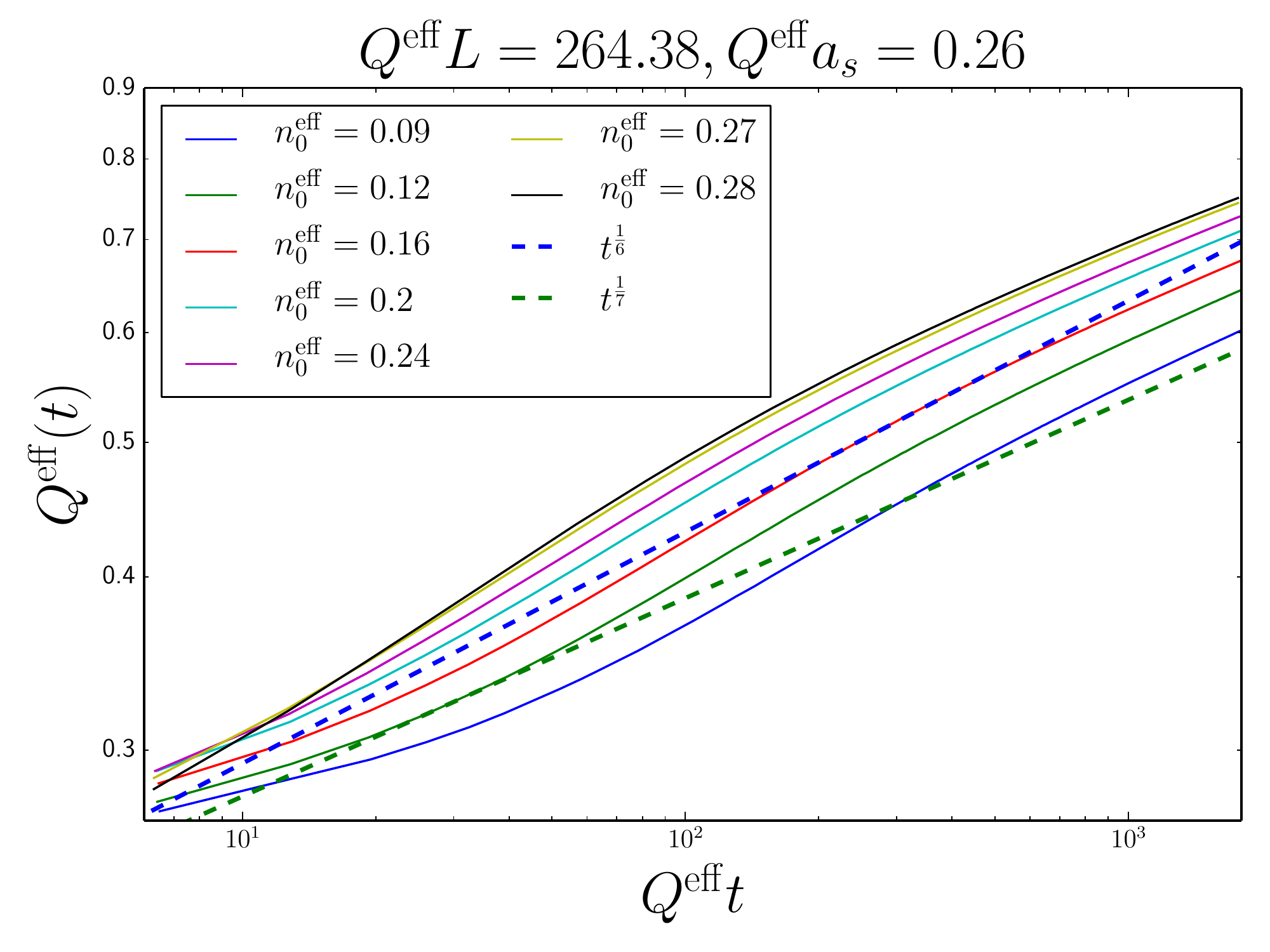}}
\caption{Time dependence of the effective momentum scale for various $n_0^{\mathrm{eff}}$. When averaged over long times, it seems that the effective momentum scale  approximately follows the $t^{\nicefrac{1}{6}}$ power law.  However, when looking at only the asymptotic regime, i.e. $tQ^{\mathrm{eff}} > 500$ it seems that $Q^{\mathrm{eff}}$ grows faster than $t^{\nicefrac{1}{7}}$, but also definitely slower than $t^{\nicefrac{1}{6}}$. The results have been averaged over five runs.}
\label{fig:Qeffvst}
\end{figure}

Both of these effective scales $Q_{\mathrm{eff}}(t)$ and $n_0^{\mathrm{eff}}(t)$ are functions of time. 
The time dependence of the momentum scale is shown in \fig\ref{fig:Qeffvst}.
In practice we find that these scales evolve in time as $Q_{\mathrm{eff}}(t) \sim t^{\nicefrac{1}{7}}-t^{\nicefrac{1}{6}}$ and consequently $n_0^{\mathrm{eff}}(t) \sim  t^{\nicefrac{-3}{7}}-t^{\nicefrac{-1}{2}}$ in two spatial dimensions.
From now on we will be using the notation $Q_{\mathrm{eff}}(t=0)=Q_{\mathrm{eff}}$ and $n_0^{\mathrm{eff}}(t=0) = n_0^{\mathrm{eff}}$.

\section{Methods for extracting the plasmon mass}\label{sec:omegapl}
We use the same methods to extract the plasmon mass as in our previous work~\cite{Lappi:2016ato}, and refer the reader there for more details.

\subsection{Uniform electric field}

\begin{figure}
\centerline{\includegraphics[width=0.48\textwidth]{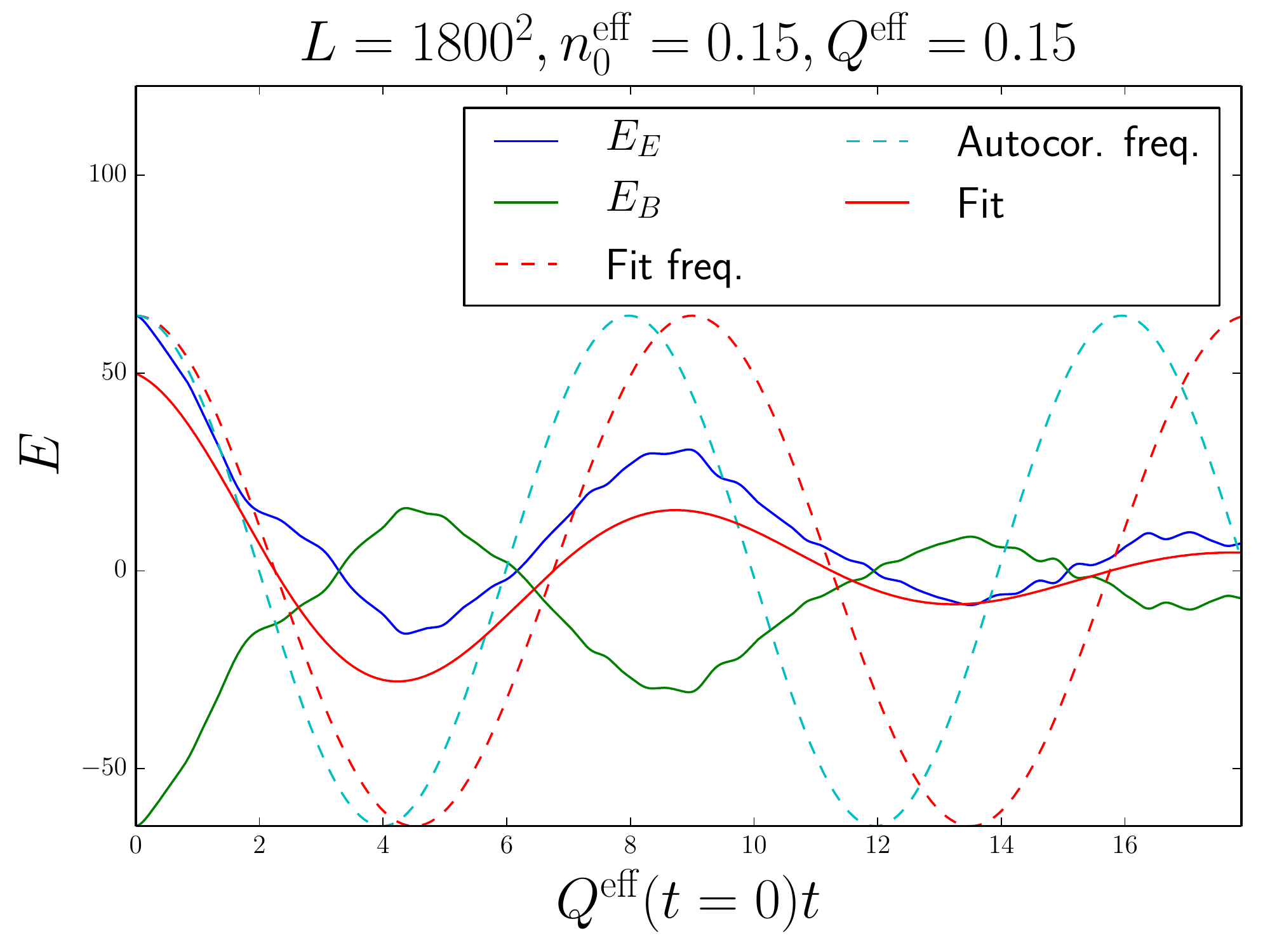}}
 \caption{Oscillation between the electric ($E_E$) and magnetic ($E_B$) energies after the addition of the homogeneous chromoelectric field. Both fields have been shifted to oscillate around 0 by subtracting the time average after the addition of the homogeneous chromoelectric field. The curve labeled as ``fit'' shows the result of the damped oscillation fit. The curves ''Fit freq.'' and ''Autocor. freq."  show the oscillations with the frequencies extracted from the fit and autocorrelation function without any damping. The fitting region  here is constrained to the part of the oscillation visible in this plot. The uniform electric field was introduced at $Q^{\mathrm{eff}}t = 160.$
}
 \label{fig:ueosc}
\end{figure}

\begin{figure}
\centerline{\includegraphics[width=0.48\textwidth]{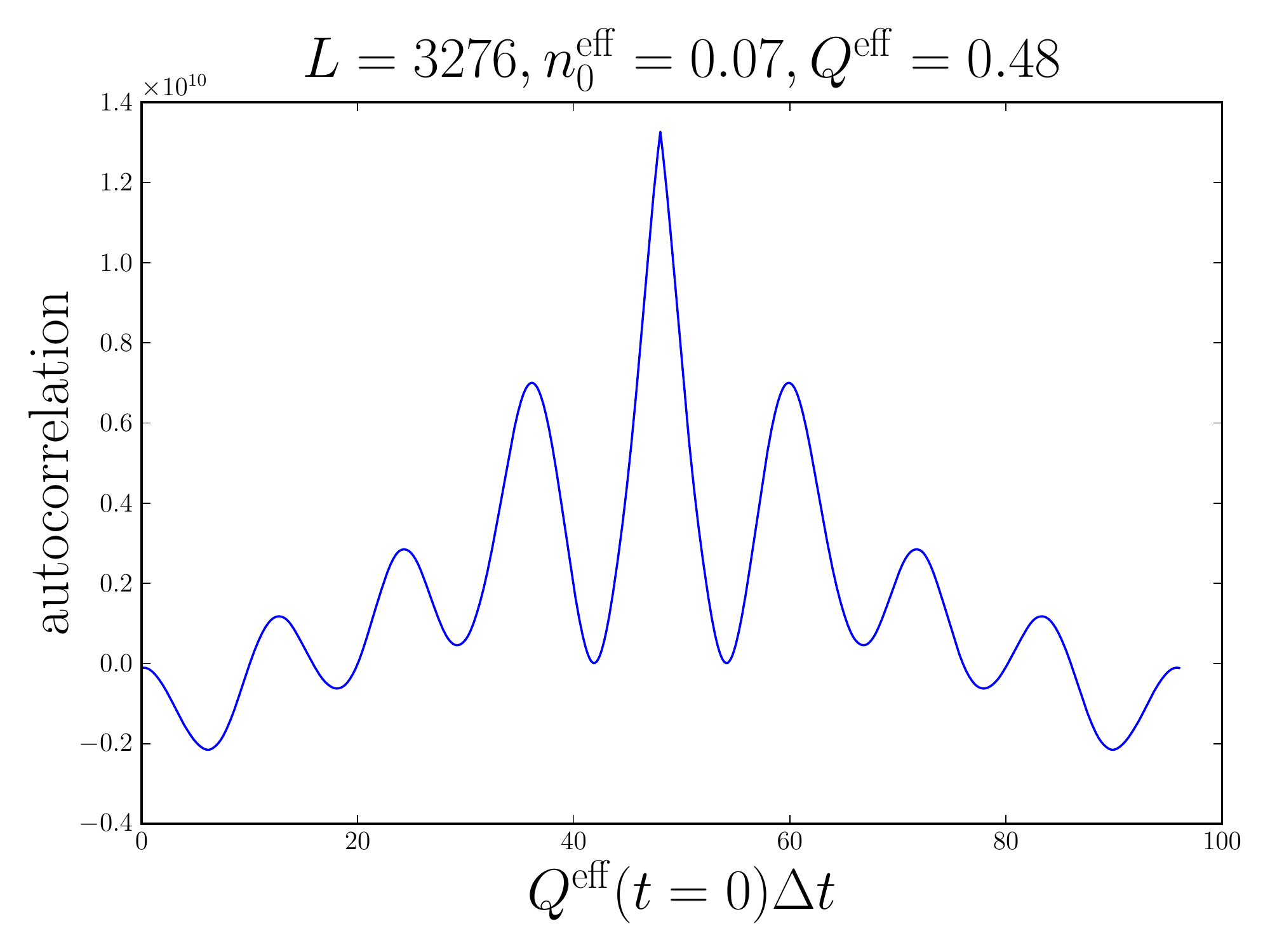}}
 \caption{Autocorrelation function of the electric energy. Due to our definition of the correlation function the peak at $Q^{\mathrm{eff}} \Delta t = 50$ corresponds to the autocorrelation of the electric field without any lag. The distance between this peak and the first maxima gives the period of the oscillation divided by 2. 
}
 \label{fig:autocor}
\end{figure}

In the three-dimensional case  it turned out~\cite{Lappi:2016ato}  that the best method to extract the plasmon mass is probably the uniform electric field method \cite{Kurkela:2012hp}. In this measurement one introduces, by hand, a uniform chromoelectric field, corresponding to  a perturbation with zero momentum. The response to this perturbation is measured in the total energy of the system, and the plasmon mass can be read off from the frequency of the oscillation between electric and magnetic energies. This procedure explicitly breaks Gauss's law, and one has to restore it by hand. The restoration is done using the algorithm described in Ref.~\cite{Moore:1996qs}.

We use two methods to measure the oscillation frequency. The first method is to fit a damped oscillation to the signal, as we have done in \fig \ref{fig:ueosc}. Note that we subtract the time average of the energy in order to move the signal to oscillate around zero. 
In practice the oscillation gets damped quite quickly and the whole time interval until the end of the simulation is  dominated by noise. To treat this situation we perform two successive fits. First by fitting to all the available data we obtain a first estimate of the oscillation period. We then perform a second fit to a time interval of two periods of the first oscillation.
The second fit gives the actual estimate for the frequency.
The electric energy is proportional to the electric field squared, and thus two oscillation periods  correspond to one oscillation in the electric field itself.  In this way we could, in principle, also extract the damping rate. In practice we have found that this method systematically overestimates the damping rate, and the fitting procedure itself does not work as well in two dimensions as in three dimensions.

The second variant of the UE  method is to compute the autocorrelation function of the electric energy and look at the separation of the maxima. We define the correlation function as 
\begin{equation}
c_{av}[k] = \sum_n a[n+k] v[n]^*,
\end{equation}
where $a$ and $v$ are sequences. If the the span of our data set in time is $\Delta t$ the correlation is computed from $k = \frac{-\Delta t}{2}$ to $k = \frac{\Delta t}{2}$. The sequences are padded with zeros whenever necessary to keep the sum well defined. When the correlation function is defined in this way, we expect the highest peak to appear, when $k$ is half of the length of the sequences. When computed in this way, the autocorrelation function is also symmetric. An example of the autocorrelation function for the electric energy is shown in \fig \ref{fig:autocor}. Here we have subtracted the time average of the energies.  At $Q^{\mathrm{eff}}\Delta t = 50$ we have the correlation of the signal with itself without any lag. 
The period of the oscillation can (remembering that the signal is  the square of the electric field)  be extracted by looking at the distance between the peak at $Q^{\mathrm{eff}} \Delta t = 50$ and the first maximum.

Note that in \fig\ref{fig:ueosc}, the actual frequency of the oscillation seems to be located between the two methods. Mathematically we would expect the damped oscillation fit to give better results than the autocorrelation method. The reason for this is that damping also shifts the location of the maxima and minima from the values given by the oscillation frequency. The autocorrelation method, in contrast, really looks for the peaks and dips of the oscillation.

\subsection{Dispersion relation} \label{subsec:DR}
\begin{figure}[]
\centerline{\includegraphics[width=0.48\textwidth]{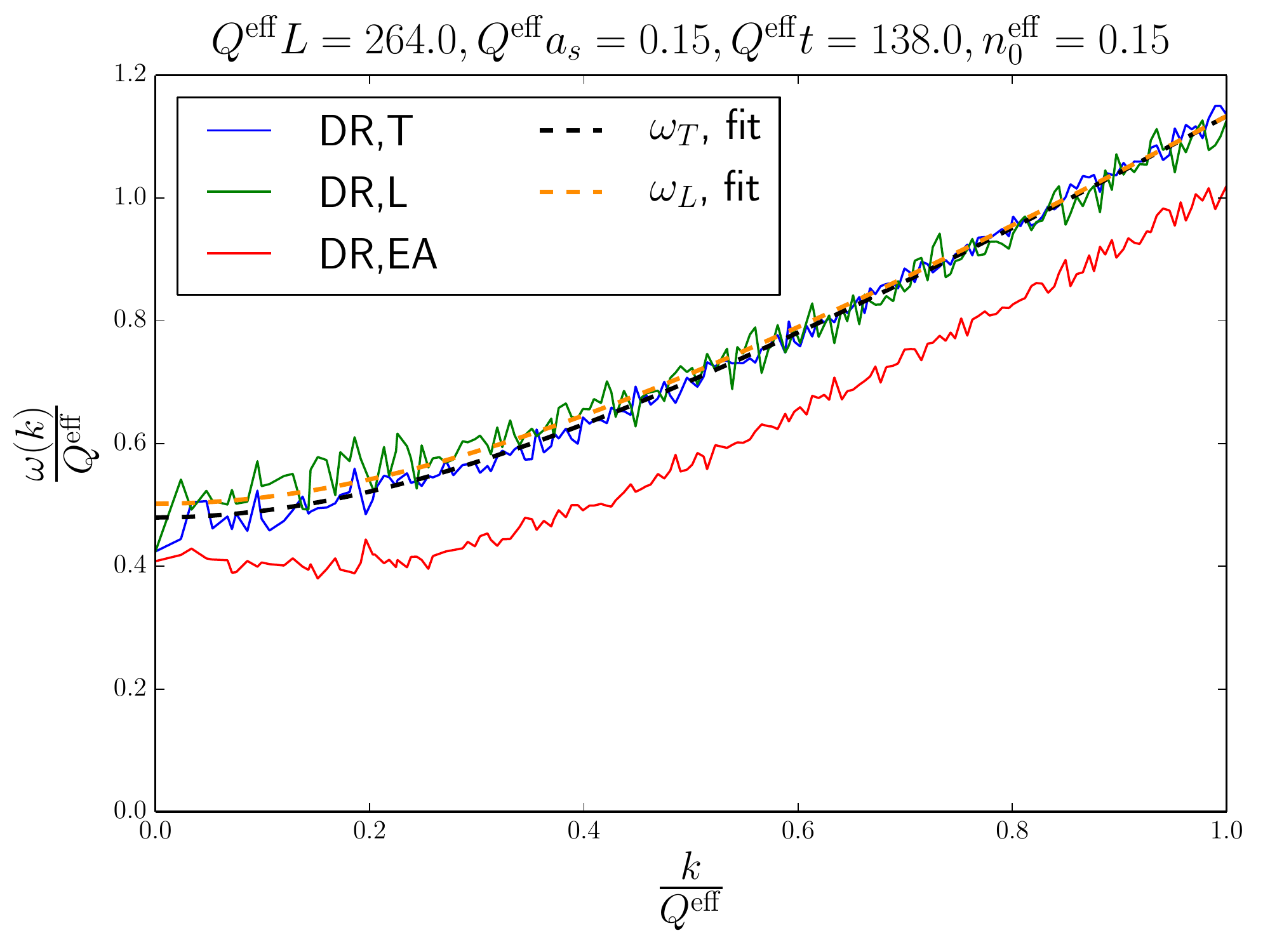}}
 \caption{Numerically extracted longitudinal (L) and transverse (T) dispersion relations and fits to these.   The maximum momentum used in the fit is $\nicefrac{k}{Q^{\mathrm{eff}}} < 0.25$ here. Averaged over 20 configurations.}
 \label{fig:DR}
\end{figure}

We extract an effective  dispersion relation from the electric field and its time derivative as
\begin{equation}
\omega^2_{T,L}\left(k\right) = \dfrac{\left<\left|\dot{E}_{i,T,L}^a\left(k\right)\right|^2 \right>}{\left< \left|E_{i,T,L}^a\left(k\right)\right|^2 \right>},
\label{eq:dispersiorelaatio}
\end{equation}
allowing us to separately study the transverse (T) and longitudinal (L) polarization. As we did for three dimensions, one can also compare the electric field and the gauge potential:
\begin{equation}
\omega^2\left(k\right) = \dfrac{\left<\left|E_{i}^a\left(k\right)\right|^2 \right>}{\left< \left|A_{i}^a\left(k\right)\right|^2 \right>},
\label{eq:huonodispersiorelaatio}
\end{equation}
as was done in \cite{Krasnitz:2000gz} for (2+1)-dimensional gauge theory. In the case of three dimensions we found~\cite{Lappi:2016ato} that \eq\nr{eq:huonodispersiorelaatio} significantly underestimates the plasmon mass. 
\begin{figure}[]
\centerline{\includegraphics[width=0.48\textwidth]{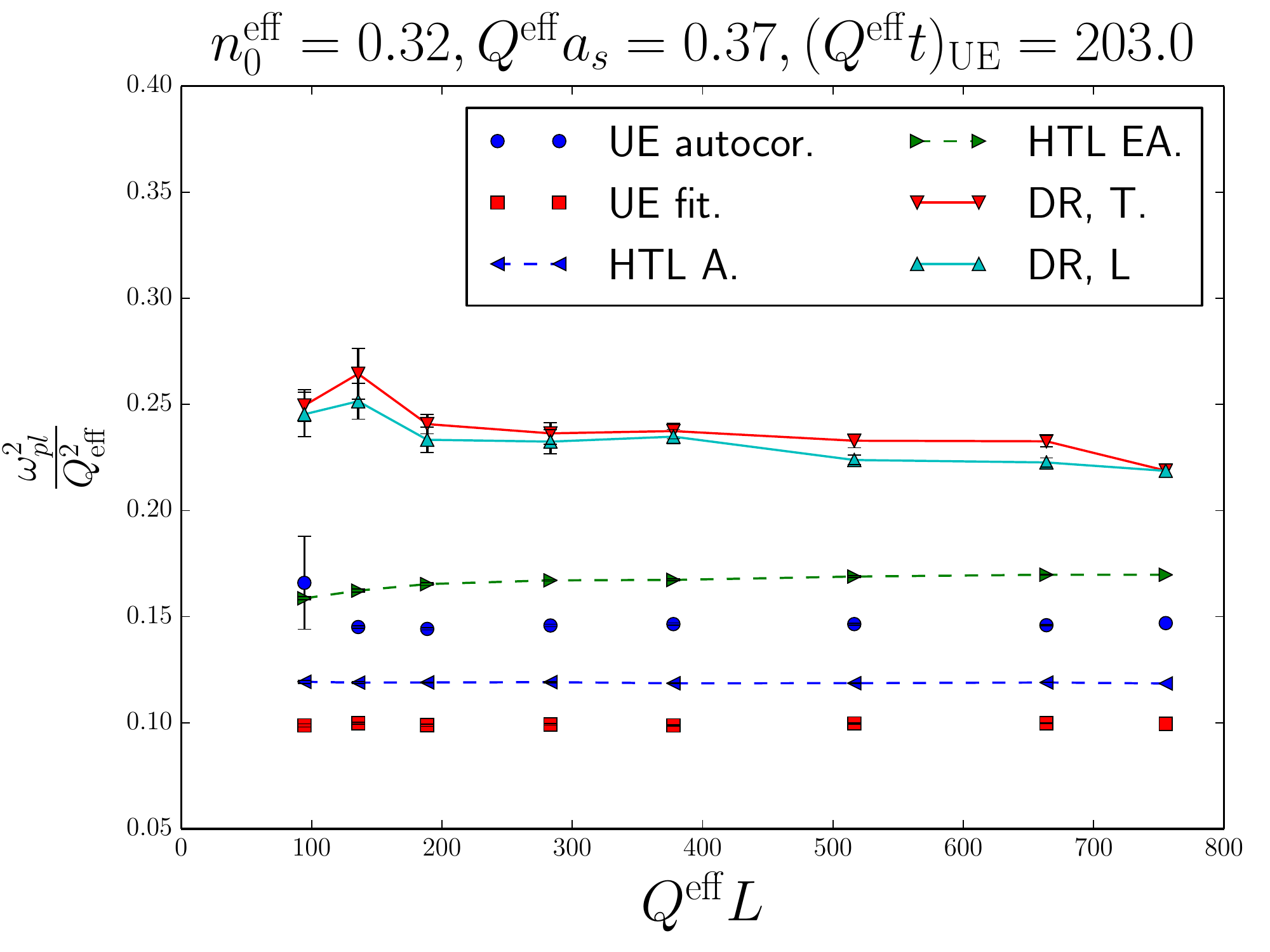}}
 \caption{Infrared cutoff dependence of the plasmon mass for various methods. The transverse and longitudinal dispersion relations extracted as in \eq (\ref{eq:dispersiorelaatio}) are denoted by 
 DR,T and DR,L. The autocorrelation and UE fit results are denoted by ``UE autocor.'' and ``UE fit''. The curves labeled HTL A and HTL EA refer to the HTL method defined by eq. (\ref{eq:htlintegral}) with occupation number extracted using \eqs (\ref{eq:discFA}) and (\ref{eq:discFEA}) respectively. Here $n_0^{\mathrm{eff}}$ varies between $n_0^{\mathrm{eff}} = 0.3240-0.3246,$ and  $Q^{\mathrm{eff}} = 0.3686-0.369,$ and the lattice sizes used were 256, 368, 512, 768, 1024, 1400, 1800, 2048, and results were averaged over 20, 15, 15, 9, 10, 8, 6, 5 configurations. The title of the plot shows the average values of  $Q^{\mathrm{eff}}$ and $n_0^{\mathrm{eff}}$. 
 }
 \label{fig:ircutoff}
\end{figure}
Typical examples of the dispersion relations \nr{eq:dispersiorelaatio} and \nr{eq:huonodispersiorelaatio} are shown in \fig\ref{fig:DR}. We also see here that the method using the fields \eq\nr{eq:huonodispersiorelaatio} gives a lower value for the mass gap. The difference is, however, nowhere near as drastic as in three dimensions, and actually the values given by \nr{eq:huonodispersiorelaatio} are closer to the values given by other methods in two dimensions. Perhaps more importantly, the evidence for a mass gap with the estimator \nr{eq:huonodispersiorelaatio} vanishes at larger momenta, while for \eq\nr{eq:dispersiorelaatio} the value remains consistent with the gap at zero momentum. Also, since in Coulomb gauge the gauge potential $A_i$ has no longitudinal component, but the magnetic part of the longitudinal plasmon is hidden in the nonlinear terms, we cannot study longitudinal modes separately with \nr{eq:huonodispersiorelaatio}. Thus, while \nr{eq:huonodispersiorelaatio} gives much more reasonable results in two dimensions than it does in three, we still consider the estimate~\nr{eq:dispersiorelaatio} using the time derivative of the electric field a better one, and will use it in the following.

The extraction of the plasmon mass is done by using a linear fit  of the  form $\omega^2 = \omega_{pl}^2 + a k^2$ (with two free parameters $\omega_{pl}^2$ and $a$) to \eq \nr{eq:dispersiorelaatio}  
The maximum momentum we use in the fit is $\nicefrac{k}{Q^{\mathrm{eff}}} < 0.25$. This cutoff has been chosen by experimenting with the fitting procedure. Choosing a significantly larger cutoff results in the fit overshooting the constant parameter. However, one still wants to have plenty of statistics for the fit, which is why one should not choose a cutoff that is too small.

\subsection{HTL resummed approximation}
\begin{figure}[]
\centerline{\includegraphics[width=0.48\textwidth]{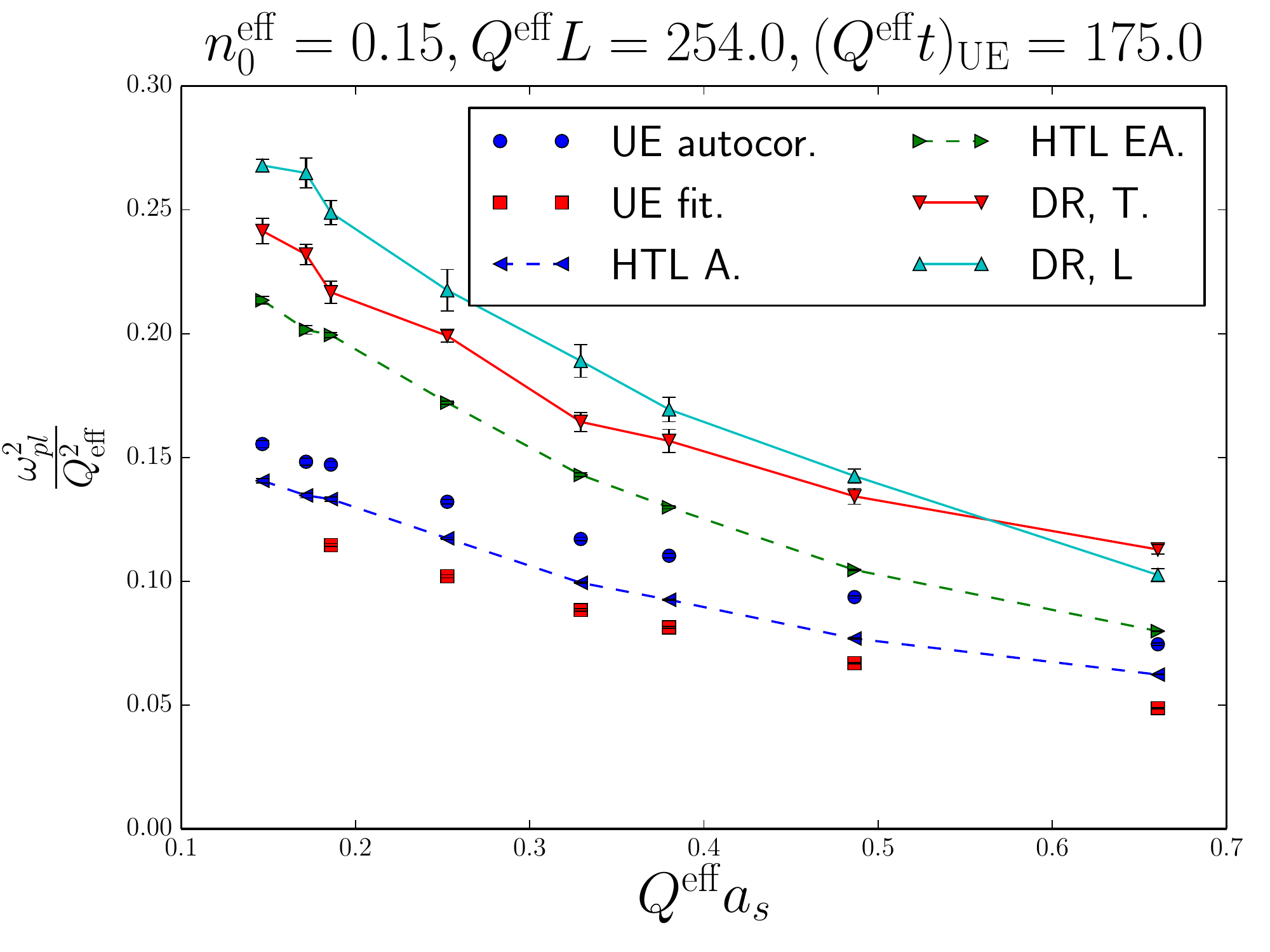}}
 \caption{The UV-cutoff dependence of the plasmon mass, with labels as in \fig\ref{fig:ircutoff}. Here $QL$ ranges from 246 to 270 and $n_0^{\mathrm{eff}}$ ranges from 0.147 to 0.162. We observe that the results given by all methods increase when we approach continuum limit. The  lattice sizes we used were $376^2, 514^2, 646^2, 763^2, 1008^2, 1326^2, 1575^2, 1800^2$ with averages taken over 10, 9, 8, 7, 6, 5, 4, 3 simulations. }
 \label{fig:uvcutoff}
\end{figure}
If the HTL  kind of separation of scales is valid, the plasmon mass is given by the integral 
\begin{equation} \label{eq:htlintegral}
\omega_{pl}^2 = \dfrac{4}{3} g^2 N_c \int \dfrac{\mathrm{d}^3k}{\left(2 \pi \right)^3} \dfrac{f\left(k \right)}{k}.
\end{equation}
On the lattice the integral is discretized by the standard replacement 
\begin{equation}
\int\dfrac{ \mathrm{d}^3k}{\left(2\pi \right)^3} \rightarrow \sum_k\dfrac{1}{V},
\end{equation}
where $k$ runs over the modes available on the lattice. This method is also widely used in the literature, see, e.g., \res \cite{Epelbaum:2011pc,Berges:2013fga,Mace:2016svc}.
While estimating the plasmon mass scale using \eq (\ref{eq:htlintegral}) one can use different definitions for the particle distribution, as discussed in Sec. \ref{sec:partdist}. As we saw from \fig \ref{fig:fmethods}, the occupation number given by the estimators $f_E$ or $f_{A+E}$ is IR divergent. Because of the different phase space, this effect is much stronger in two dimensions than it is in three. In fact, an infrared  convergent value for the integral \nr{eq:htlintegral} with the estimators $f_{A+E}$ or $f_E$ would require the $EE$ correlator to approach zero for $k\to 0$. This would be a large suppression compared to the thermal value $|E(k)|^2 \sim T$, and would, in a contradictory fashion, correspond to a vanishing mass gap in the dispersion relation \nr{eq:huonodispersiorelaatio}.
Thus we use only the estimators $f_A$ and $f_{EA}$ that yield a finite value to compute the plasmon mass using \eq (\ref{eq:htlintegral}). We refer to these as HTL-A and HTL-EA.

\section{Dependence on lattice cutoffs, time and occupation number} \label{sec:results}

\begin{figure}[]
\centerline{\includegraphics[width=0.48\textwidth]
{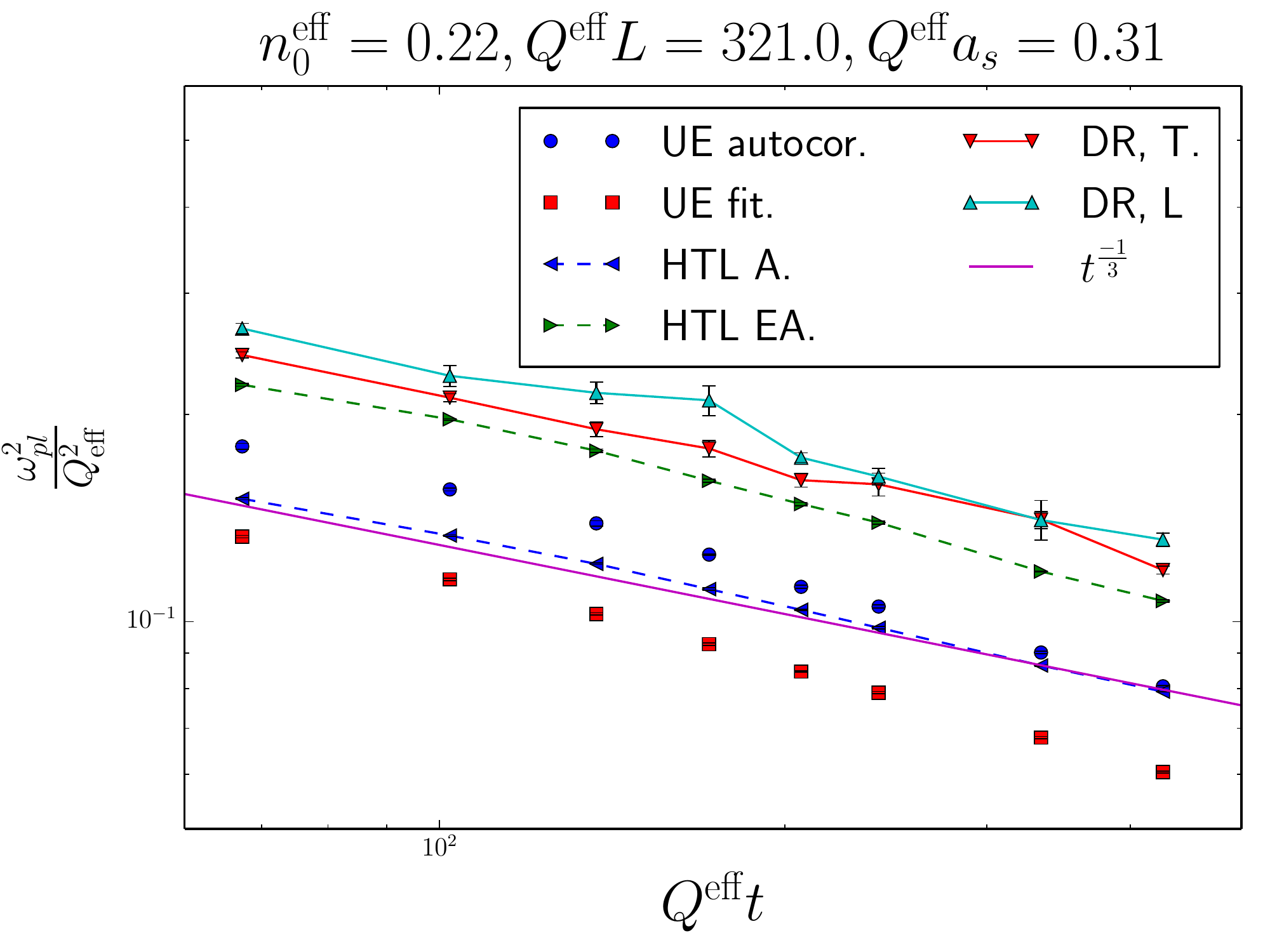}}
\caption{Time dependence of the plasmon mass scale with various methods, with labels as in \fig\ref{fig:ircutoff}. We also show a curve corresponding to $t^{\nicefrac{-1}{3}}$ power law. We find that the HTL-A method almost agrees with this power law.  It seems that the value given by the UE measurement decreases slightly faster than that of HTL methods and the power law. Results are averages over five runs. }\label{fig:tdep3}
\end{figure}

Next we  study how our results depend on lattice cutoffs. The ultraviolet cutoff is set by $Q^{\mathrm{eff}}a_s$ and the infrared cutoff is controlled by $Q^{\mathrm{eff}}L$. In practice when we study the dependencies we vary one of these cutoffs and keep the other fixed. The fact that we are using the gauge invariant observables $Q^{\mathrm{eff}}$ and $n_0^{\mathrm{eff}}$ makes the choice of initial parameters $Q$ and $n_0$ rather complicated. The necessity to invert the relation shown in  \figs \ref{fig:QeffvsQ} and \ref{fig:n0effvsQ} to choose $Q$ and $n_0$ corresponding to fixed $Q^{\mathrm{eff}}$ and $n_0^{\mathrm{eff}}$ introduces some additional uncertainty into these estimates. 
The error bars shown in the figures in this section are statistical errors computed as the standard error of the mean. It turns out that these errors are insignificant for HTL and UE methods, but for the DR method  the statistical errors are larger.
The infrared cutoff dependence is shown in \fig \ref{fig:ircutoff}. All observables seem to be well behaved in the infinite volume limit. Our results on the ultraviolet cutoff dependence are shown in \fig \ref{fig:uvcutoff}. A remarkable feature is that values given by all methods seem to increase when we approach the continuum limit, in contrast to the behavior observed in the three-dimensional system~\cite{Lappi:2016ato}. However the results do not seem to be UV divergent. 

\begin{figure}[]
\centerline{\includegraphics[width=0.48\textwidth]
{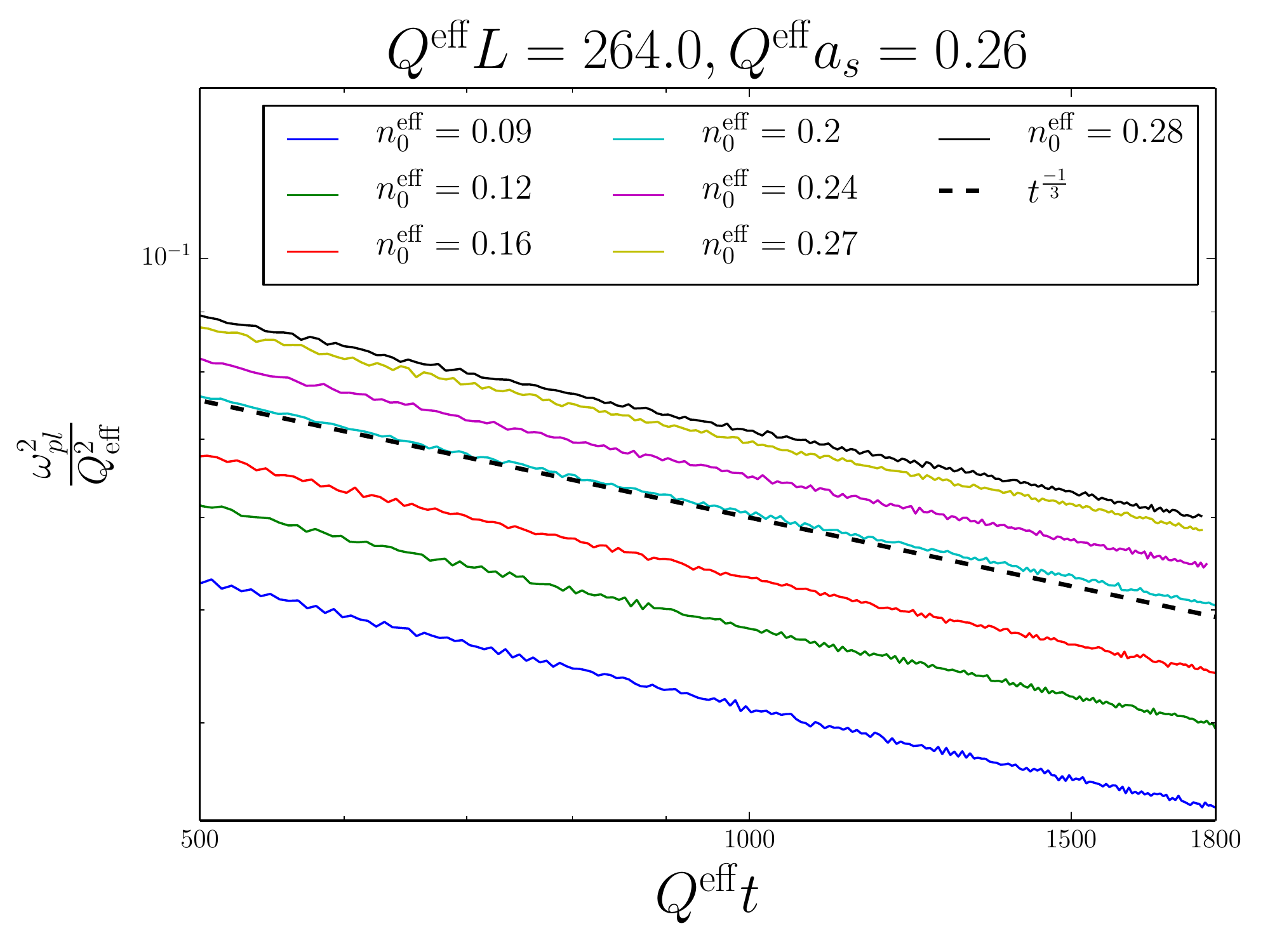}}
\caption{Time dependence using only the HTL-A method. We also show the $t^{\nicefrac{-1}{3}}$ power law here for comparison. It seems that at late times the HTL-A method agrees with the power law. The results are  averaged over five runs.}
\label{fig:HTLAvst}
\end{figure}

The time dependence of the plasmon mass using various methods at early times is shown in \fig \ref{fig:tdep3}. We find that the time evolution computed with the HTL-A method agrees reasonably well with  a $t^{\nicefrac{-1}{3}}$ power law. However it seems that the UE method decreases slightly faster than the HTL-A method and the power law. We also find that the difference between DR and the other methods persists even at late times. In order to study the asymptotic power law behaviour on longer timescales we do a study using only the HTL-A method (using the UE method would require prohibitively many separate runs).  The results are shown in \fig \ref{fig:HTLAvst}. We find that at late times the HTL-A method is in agreement with the proposed power law. 

\begin{figure}[]
\centerline{\includegraphics[width=0.48\textwidth]{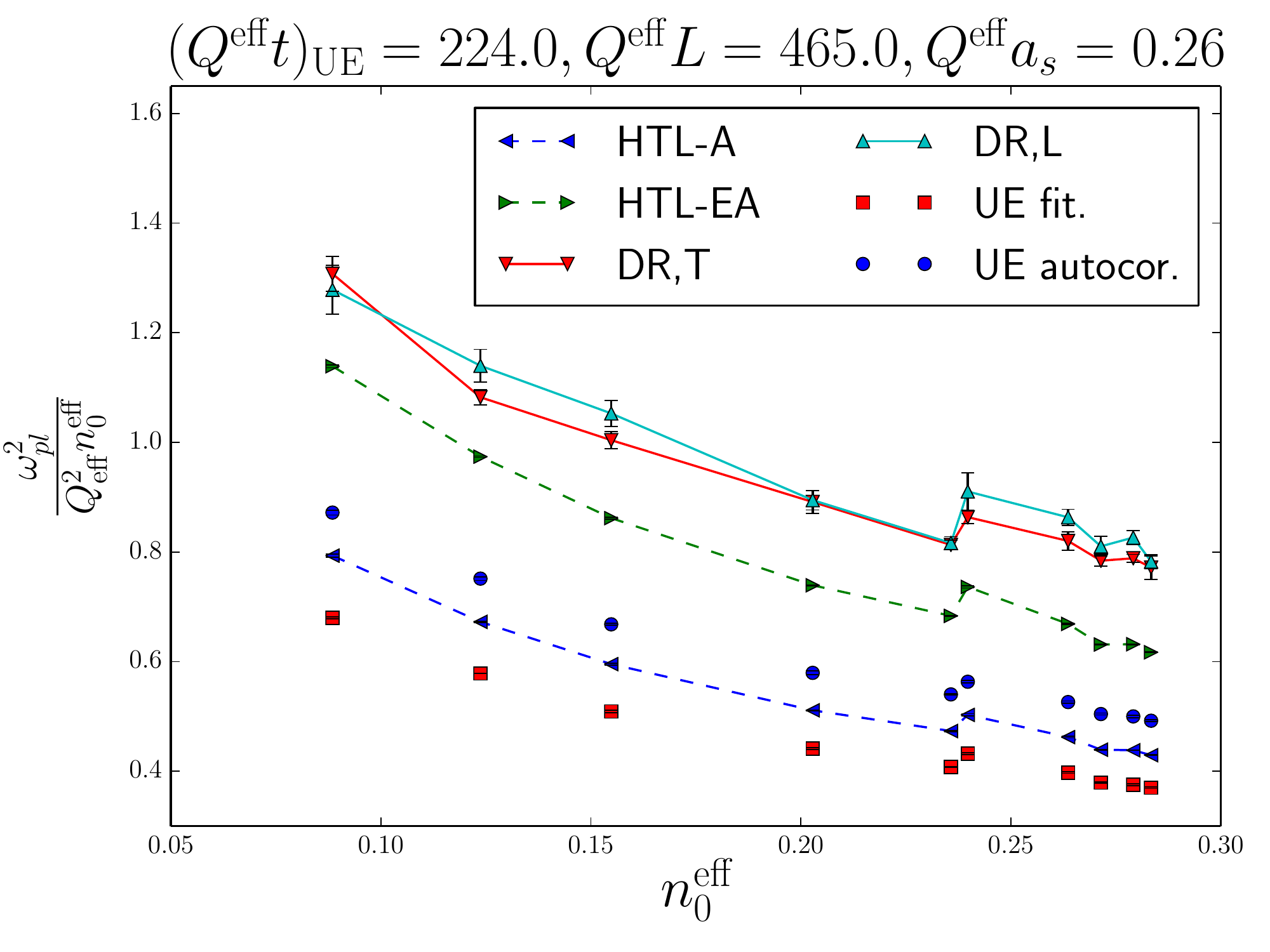}}
 \caption{Dependence of plasmon mass (scaled by the occupation number $n_0^{\mathrm{eff}}$) on the occupation number $n_0^{\mathrm{eff}}$ for the different methods of evaluating the plasmon mass scale at fixed time, with labels as in \fig\ref{fig:ircutoff}. We observe a similar trend as we did in three dimensions: the plasmon mass scale at a fixed time decreases as a function of the occupation number.  Here $Q^{\mathrm{eff}}L=432-480.$ and $Q^{\mathrm{eff}} = 0.240-0.267.$ $(Q^{\mathrm{eff}}t)_\mathrm{UE} = 205-227.$ The bumb at $n_0 \approx 0.24$ is caused by a deviation in these parameters. These results are averaged over five configurations.}
\label{fig:occupnumberdep}
\end{figure}

The occupation number dependence is depicted in \fig \ref{fig:occupnumberdep}. The dependence seems qualitatively similar to the three dimensional case - the plasmon mass scale (normalized by the occupation number) falls as  function of increasing $n_0^{\mathrm{eff}}$.  The bump in the figure roughly at $n_0^{\mathrm{eff}} = 0.24$ is caused by deviation in the input parameters. It also seems that the differences between the methods are independent of occupation number. This is suggestive of a similar interpretation as the three-dimensional results~\cite{Lappi:2016ato}, namely that larger occupation numbers lead to a more rapid start of the large time scaling regime where the plasmon mass begins to fall with time.

\section{Conclusions and outlook}\label{sec:conc}
In this paper we have first argued for a need for gauge invariant observables to measure the occupation number and momentum scale. This need arose due to gauge fixing effects, which turned out to deform the spectrum more than one might have expected based on our previous three dimensional simulations.

According to our observations the UE and HTL-A methods are in rough agreement, and they are almost equally good in measuring the plasmon mass scale. However, the HTL-A method is computationally much cheaper and easier to implement, so we recommend that the reader use it when measuring the plasmon mass scale in two-dimensional simulations. However, one should bear in mind that in two spatial dimensions the dependence on the precise definition of the quasiparticle distribution is much larger than in three, because the HTL integral is more dominated by the infrared. The DR method works in a similar fashion as in the three dimensional case - the agreement with the other methods is within a factor of 2. 
 
We have also studied the cutoff dependencies of our results. We find no cutoff dependencies when we vary the infrared cutoff. However the choice of the ultraviolet cutoff does influence the results we obtain. It turns out that the values given by all methods increase when we get closer to the continuum. This behavior is different from the three dimensional case, where we observed that in the continuum limit the agreement between the UE and HTL results seemed to improve. We find no evidence of this in two spatial dimensions. 

The dependence on the occupation number is similar to the behaviour in three dimensions, and the differences between the methods persist regardless of occupation number. We observe that the plasmon mass scale squared seems to decrease like $t^{\nicefrac{-1}{3}}$ at late times when measured with the HTL-A method. We also observe a reasonably good agreement at earlier times. However at earlier times the UE measurement seems to decrease slightly faster than this power law.

It would be interesting to study whether two-dimensional classical Yang-Mills system can be understood in a kinetic theory framework. However, in two dimensions the contribution of the modes below the Debye scale becomes a lot more important than in three dimensions, which we can see from the deviations between the different HTL estimates.

We are planning to address the dispersion relation and spectral function in classical Yang-Mills systems in future publication using our recently developed techniques to perform simulations with fluctuations on top of the classical background \cite{Kurkela:2016mhu}. In this way we could also address the existence of quasiparticles in classical Yang-Mills theory, which would shed light on the kinetic theory description of classical Yang-Mills systems.

\begin{acknowledgments}
  We are grateful to A. Kurkela for discussions.
  T.~L.\ is supported by the Academy of Finland, projects No. 267321 and No. 303756  and by the European Research Council, grant ERC-2015-CoG-681707. J.P. is supported by the Jenny and Antti Wihuri Foundation. This work has used computational resources from CSC – IT Center for Science, Finland.
\end{acknowledgments}

\bibliography{spires}

\providecommand{\href}[2]{#2}\begingroup\raggedright\begin{thebibliography}{10}

\bibitem{Iancu:2003xm}
E.~Iancu and R.~Venugopalan in {\em Quark gluon plasma} (R.~Hwa and X.~N. Wang,
  eds.).
\newblock World Scientific, 2003.
\newblock \href{http://arXiv.org/abs/hep-ph/0303204}{{\tt
  arXiv:hep-ph/0303204}}.
%%CITATION = HEP-PH 0303204;%%

\bibitem{Gelis:2010nm}
F.~Gelis, E.~Iancu, J.~Jalilian-Marian and R.~Venugopalan, {\it The color glass
  condensate},
  \href{http://dx.doi.org/10.1146/annurev.nucl.010909.083629}{{\em Ann. Rev.
  Nucl. Part. Sci.} {\bf 60} (2010) 463}
  [\href{http://arXiv.org/abs/1002.0333}{{\tt arXiv:1002.0333 [hep-ph]}}].

\bibitem{Albacete:2014fwa}
J.~L. Albacete and C.~Marquet, {\it {Gluon saturation and initial conditions
  for relativistic heavy ion collisions}},
  \href{http://dx.doi.org/10.1016/j.ppnp.2014.01.004}{{\em Prog. Part. Nucl.
  Phys.} {\bf 76} (2014) 1} [\href{http://arXiv.org/abs/1401.4866}{{\tt
  arXiv:1401.4866 [hep-ph]}}].
%%CITATION = ARXIV:1401.4866;%%

\bibitem{Lappi:2003bi}
T.~Lappi, {\it Production of gluons in the classical field model for heavy ion
  collisions},  \href{http://dx.doi.org/10.1103/PhysRevC.67.054903}{{\em Phys.
  Rev.} {\bf C67} (2003) 054903}
  [\href{http://arXiv.org/abs/hep-ph/0303076}{{\tt arXiv:hep-ph/0303076}}].
%%CITATION = HEP-PH/0303076;%%

\bibitem{Mueller:2002gd}
A.~H. Mueller and D.~T. Son, {\it On the equivalence between the boltzmann
  equation and classical field theory at large occupation numbers},  {\em Phys.
  Lett.} {\bf B582} (2004) 279 [\href{http://arXiv.org/abs/hep-ph/0212198}{{\tt
  arXiv:hep-ph/0212198}}].
%%CITATION = HEP-PH 0212198;%%

\bibitem{Jeon:2004dh}
S.~Jeon, {\it The boltzmann equation in classical and quantum field theory},
  {\em Phys. Rev.} {\bf C72} (2005) 014907
  [\href{http://arXiv.org/abs/hep-ph/0412121}{{\tt arXiv:hep-ph/0412121}}].
%%CITATION = HEP-PH 0412121;%%

\bibitem{Berges:2004yj}
J.~Berges, {\it Introduction to nonequilibrium quantum field theory},
  \href{http://arXiv.org/abs/hep-ph/0409233}{{\tt arXiv:hep-ph/0409233}}.
%%CITATION = HEP-PH 0409233;%%

\bibitem{Kurkela:2015qoa}
A.~Kurkela and Y.~Zhu, {\it Isotropization and hydrodynamization in weakly
  coupled heavy-ion collisions},
  \href{http://dx.doi.org/10.1103/PhysRevLett.115.182301}{{\em Phys. Rev.
  Lett.} {\bf 115} (2015) 182301} [\href{http://arXiv.org/abs/1506.06647}{{\tt
  arXiv:1506.06647 [hep-ph]}}].
%%CITATION = ARXIV:1506.06647;%%

\bibitem{Krasnitz:1998ns}
A.~Krasnitz and R.~Venugopalan, {\it Non-perturbative computation of gluon
  mini-jet production in nuclear collisions at very high energies},
  \href{http://dx.doi.org/10.1016/S0550-3213(99)00366-1}{{\em Nucl. Phys.} {\bf
  B557} (1999) 237} [\href{http://arXiv.org/abs/hep-ph/9809433}{{\tt
  arXiv:hep-ph/9809433}}].
%%CITATION = HEP-PH/9809433;%%

\bibitem{Lappi:2009xa}
T.~Lappi, S.~Srednyak and R.~Venugopalan, {\it Non-perturbative computation of
  double inclusive gluon production in the glasma},
  \href{http://dx.doi.org/10.1007/JHEP01(2010)066}{{\em JHEP} {\bf 1001} (2010)
  066} [\href{http://arXiv.org/abs/0911.2068}{{\tt arXiv:0911.2068 [hep-ph]}}].
%%CITATION = 0911.2068;%%

\bibitem{Kovner:1995ts}
A.~Kovner, L.~D. McLerran and H.~Weigert, {\it Gluon production at high
  transverse momentum in the mclerran-venugopalan model of nuclear structure
  functions},  \href{http://dx.doi.org/10.1103/PhysRevD.52.3809}{{\em Phys.
  Rev.} {\bf D52} (1995) 3809} [\href{http://arXiv.org/abs/hep-ph/9505320}{{\tt
  arXiv:hep-ph/9505320}}].
%%CITATION = HEP-PH/9505320;%%

\bibitem{Kovner:1995ja}
A.~Kovner, L.~D. McLerran and H.~Weigert, {\it Gluon production from
  non-abelian {Weizs{\"a}cker}-{Williams} fields in nucleus-nucleus
  collisions},  \href{http://dx.doi.org/10.1103/PhysRevD.52.6231}{{\em Phys.
  Rev.} {\bf D52} (1995) 6231} [\href{http://arXiv.org/abs/hep-ph/9502289}{{\tt
  arXiv:hep-ph/9502289}}].
%%CITATION = HEP-PH/9502289;%%

\bibitem{Lappi:2006fp}
T.~Lappi and L.~McLerran, {\it Some features of the glasma},
  \href{http://dx.doi.org/10.1016/j.nuclphysa.2006.04.001}{{\em Nucl. Phys.}
  {\bf A772} (2006) 200} [\href{http://arXiv.org/abs/hep-ph/0602189}{{\tt
  arXiv:hep-ph/0602189}}].
%%CITATION = HEP-PH/0602189;%%

\bibitem{Gelfand:2016yho}
D.~Gelfand, A.~Ipp and D.~M{\"u}ller, {\it Simulating collisions of thick
  nuclei in the color glass condensate framework},
  \href{http://dx.doi.org/10.1103/PhysRevD.94.014020}{{\em Phys. Rev.} {\bf
  D94} (2016) 014020} [\href{http://arXiv.org/abs/1605.07184}{{\tt
  arXiv:1605.07184 [hep-ph]}}].
%%CITATION = ARXIV:1605.07184;%%

\bibitem{Schenke:2016ksl}
B.~Schenke and S.~Schlichting, {\it {3D glasma initial state for relativistic
  heavy ion collisions}},
  \href{http://dx.doi.org/10.1103/PhysRevC.94.044907}{{\em Phys. Rev.} {\bf
  C94} (2016)~no.~4 044907} [\href{http://arXiv.org/abs/1605.07158}{{\tt
  arXiv:1605.07158 [hep-ph]}}].
%%CITATION = ARXIV:1605.07158;%%

\bibitem{Ipp:2017lho}
A.~Ipp and D.~M{\"u}ller, {\it {Broken boost invariance in the Glasma via
  finite nuclei thickness}},
  \href{http://dx.doi.org/10.1016/j.physletb.2017.05.032}{{\em Phys. Lett.}
  {\bf B771} (2017) 74} [\href{http://arXiv.org/abs/1703.00017}{{\tt
  arXiv:1703.00017 [hep-ph]}}].
%%CITATION = ARXIV:1703.00017;%%

\bibitem{Fukushima:2006ax}
K.~Fukushima, F.~Gelis and L.~McLerran, {\it Initial singularity of the little
  bang},  \href{http://dx.doi.org/10.1016/j.nuclphysa.2007.01.086}{{\em Nucl.
  Phys.} {\bf A786} (2007) 107}
  [\href{http://arXiv.org/abs/hep-ph/0610416}{{\tt arXiv:hep-ph/0610416}}].
%%CITATION = HEP-PH/0610416;%%

\bibitem{Dusling:2010rm}
K.~Dusling, T.~Epelbaum, F.~Gelis and R.~Venugopalan, {\it Role of quantum
  fluctuations in a system with strong fields: Onset of hydrodynamical flow},
  \href{http://dx.doi.org/10.1016/j.nuclphysa.2010.11.009}{{\em Nucl. Phys.}
  {\bf A850} (2011) 69} [\href{http://arXiv.org/abs/1009.4363}{{\tt
  arXiv:1009.4363 [hep-ph]}}].

\bibitem{Epelbaum:2011pc}
T.~Epelbaum and F.~Gelis, {\it Role of quantum fluctuations in a system with
  strong fields: Spectral properties and thermalization},
  \href{http://dx.doi.org/10.1016/j.nuclphysa.2011.09.019}{{\em Nucl. Phys.}
  {\bf A872} (2011) 210} [\href{http://arXiv.org/abs/1107.0668}{{\tt
  arXiv:1107.0668 [hep-ph]}}].
%%CITATION = ARXIV:1107.0668;%%

\bibitem{Dusling:2012ig}
K.~Dusling, T.~Epelbaum, F.~Gelis and R.~Venugopalan, {\it {Instability induced
  pressure isotropization in a longitudinally expanding system}},
  \href{http://dx.doi.org/10.1103/PhysRevD.86.085040}{{\em Phys. Rev.} {\bf
  D86} (2012) 085040} [\href{http://arXiv.org/abs/1206.3336}{{\tt
  arXiv:1206.3336 [hep-ph]}}].
%%CITATION = ARXIV:1206.3336;%%

\bibitem{Epelbaum:2013waa}
T.~Epelbaum and F.~Gelis, {\it Fluctuations of the initial color fields in high
  energy heavy ion collisions},
  \href{http://dx.doi.org/10.1103/PhysRevD.88.085015}{{\em Phys. Rev.} {\bf
  D88} (2013) 085015} [\href{http://arXiv.org/abs/1307.1765}{{\tt
  arXiv:1307.1765 [hep-ph]}}].
%%CITATION = ARXIV:1307.1765;%%

\bibitem{Mrowczynski:1994xv}
S.~Mr{\'o}wczy{\'n}ski, {\it Color collective effects at the early stage of
  ultrarelativistic heavy ion collisions},
  \href{http://dx.doi.org/10.1103/PhysRevC.49.2191}{{\em Phys. Rev.} {\bf C49}
  (1994) 2191}.
%%CITATION = PHRVA,C49,2191;%%

\bibitem{Mrowczynski:1996vh}
S.~Mr{\'o}wczy{\'n}ski, {\it Color filamentation in ultrarelativistic heavy-ion
  collisions},  \href{http://dx.doi.org/10.1016/S0370-2693(96)01621-8}{{\em
  Phys. Lett.} {\bf B393} (1997) 26}
  [\href{http://arXiv.org/abs/hep-ph/9606442}{{\tt arXiv:hep-ph/9606442}}].
%%CITATION = HEP-PH/9606442;%%

\bibitem{Mrowczynski:2004kv}
S.~Mr{\'o}wczy{\'n}ski, A.~Rebhan and M.~Strickland, {\it Hard-loop effective
  action for anisotropic plasmas},
  \href{http://dx.doi.org/10.1103/PhysRevD.70.025004}{{\em Phys. Rev.} {\bf
  D70} (2004) 025004} [\href{http://arXiv.org/abs/hep-ph/0403256}{{\tt
  arXiv:hep-ph/0403256}}].
%%CITATION = HEP-PH/0403256;%%

\bibitem{Romatschke:2003ms}
P.~Romatschke and M.~Strickland, {\it Collective modes of an anisotropic
  quark-gluon plasma},
  \href{http://dx.doi.org/10.1103/PhysRevD.68.036004}{{\em Phys. Rev.} {\bf
  D68} (2003) 036004} [\href{http://arXiv.org/abs/hep-ph/0304092}{{\tt
  arXiv:hep-ph/0304092}}].
%%CITATION = HEP-PH/0304092;%%

\bibitem{Kurkela:2011ub}
A.~Kurkela and G.~D. Moore, {\it Bjorken flow, plasma instabilities, and
  thermalization},  \href{http://dx.doi.org/10.1007/JHEP11(2011)120}{{\em JHEP}
  {\bf 11} (2011) 120} [\href{http://arXiv.org/abs/1108.4684}{{\tt
  arXiv:1108.4684 [hep-ph]}}].
%%CITATION = ARXIV:1108.4684;%%

\bibitem{Arnold:2003rq}
P.~Arnold, J.~Lenaghan and G.~D. Moore, {\it {QCD} plasma instabilities and
  bottom-up thermalization},
  \href{http://dx.doi.org/10.1088/1126-6708/2003/08/002}{{\em JHEP} {\bf 08}
  (2003) 002} [\href{http://arXiv.org/abs/hep-ph/0307325}{{\tt
  arXiv:hep-ph/0307325}}].
%%CITATION = HEP-PH/0307325;%%

\bibitem{Romatschke:2004jh}
P.~Romatschke and M.~Strickland, {\it Collective modes of an anisotropic
  quark-gluon plasma. ii},
  \href{http://dx.doi.org/10.1103/PhysRevD.70.116006}{{\em Phys. Rev.} {\bf
  D70} (2004) 116006} [\href{http://arXiv.org/abs/hep-ph/0406188}{{\tt
  arXiv:hep-ph/0406188}}].
%%CITATION = HEP-PH/0406188;%%

\bibitem{Arnold:2004ti}
P.~B. Arnold, J.~Lenaghan, G.~D. Moore and L.~G. Yaffe, {\it Apparent
  thermalization due to plasma instabilities in quark-gluon plasma},
  \href{http://dx.doi.org/10.1103/PhysRevLett.94.072302}{{\em Phys. Rev. Lett.}
  {\bf 94} (2005) 072302} [\href{http://arXiv.org/abs/nucl-th/0409068}{{\tt
  arXiv:nucl-th/0409068 [nucl-th]}}].
%%CITATION = NUCL-TH/0409068;%%

\bibitem{Arnold:2004ih}
P.~B. Arnold and J.~Lenaghan, {\it The {Abelianization} of {QCD} plasma
  instabilities},  \href{http://dx.doi.org/10.1103/PhysRevD.70.114007}{{\em
  Phys. Rev.} {\bf D70} (2004) 114007}
  [\href{http://arXiv.org/abs/hep-ph/0408052}{{\tt arXiv:hep-ph/0408052
  [hep-ph]}}].
%%CITATION = HEP-PH/0408052;%%

\bibitem{Rebhan:2005re}
A.~Rebhan, P.~Romatschke and M.~Strickland, {\it {Dynamics of
  quark-gluon-plasma instabilities in discretized hard-loop approximation}},
  \href{http://dx.doi.org/10.1088/1126-6708/2005/09/041}{{\em JHEP} {\bf 09}
  (2005) 041} [\href{http://arXiv.org/abs/hep-ph/0505261}{{\tt
  arXiv:hep-ph/0505261 [hep-ph]}}].
%%CITATION = HEP-PH/0505261;%%

\bibitem{Rebhan:2004ur}
A.~Rebhan, P.~Romatschke and M.~Strickland, {\it {Hard-loop dynamics of
  non-Abelian plasma instabilities}},
  \href{http://dx.doi.org/10.1103/PhysRevLett.94.102303}{{\em Phys. Rev. Lett.}
  {\bf 94} (2005) 102303} [\href{http://arXiv.org/abs/hep-ph/0412016}{{\tt
  arXiv:hep-ph/0412016 [hep-ph]}}].
%%CITATION = HEP-PH/0412016;%%

\bibitem{Kurkela:2011ti}
A.~Kurkela and G.~D. Moore, {\it Thermalization in weakly coupled nonabelian
  plasmas},  \href{http://dx.doi.org/10.1007/JHEP12(2011)044}{{\em JHEP} {\bf
  12} (2011) 044} [\href{http://arXiv.org/abs/1107.5050}{{\tt arXiv:1107.5050
  [hep-ph]}}].
%%CITATION = ARXIV:1107.5050;%%

\bibitem{Nara:2005fr}
Y.~Nara, {\it Isotropization by {QCD} plasma instabilities},
  \href{http://dx.doi.org/10.1016/j.nuclphysa.2006.06.021}{{\em Nucl. Phys.}
  {\bf A774} (2006) 783} [\href{http://arXiv.org/abs/nucl-th/0509052}{{\tt
  arXiv:nucl-th/0509052}}].
%%CITATION = NUCL-TH/0509052;%%

\bibitem{Dumitru:2005gp}
A.~Dumitru and Y.~Nara, {\it {QCD} plasma instabilities and isotropization},
  \href{http://dx.doi.org/10.1016/j.physletb.2005.06.041}{{\em Phys. Lett.}
  {\bf B621} (2005) 89} [\href{http://arXiv.org/abs/hep-ph/0503121}{{\tt
  arXiv:hep-ph/0503121}}].
%%CITATION = HEP-PH/0503121;%%

\bibitem{Bodeker:2007fw}
D.~Bodeker and K.~Rummukainen, {\it Non-abelian plasma instabilities for strong
  anisotropy},  \href{http://dx.doi.org/10.1088/1126-6708/2007/07/022}{{\em
  JHEP} {\bf 07} (2007) 022} [\href{http://arXiv.org/abs/0705.0180}{{\tt
  arXiv:0705.0180 [hep-ph]}}].
%%CITATION = 0705.0180;%%

\bibitem{Rebhan:2008uj}
A.~Rebhan, M.~Strickland and M.~Attems, {\it {Instabilities of an
  anisotropically expanding non-Abelian plasma: 1D+3V discretized hard-loop
  simulations}},  \href{http://dx.doi.org/10.1103/PhysRevD.78.045023}{{\em
  Phys. Rev.} {\bf D78} (2008) 045023}
  [\href{http://arXiv.org/abs/0802.1714}{{\tt arXiv:0802.1714 [hep-ph]}}].
%%CITATION = ARXIV:0802.1714;%%

\bibitem{Attems:2012js}
M.~Attems, A.~Rebhan and M.~Strickland, {\it {Instabilities of an
  anisotropically expanding non-Abelian plasma: 3D+3V discretized hard-loop
  simulations}},  \href{http://dx.doi.org/10.1103/PhysRevD.87.025010}{{\em
  Phys. Rev.} {\bf D87} (2013)~no.~2 025010}
  [\href{http://arXiv.org/abs/1207.5795}{{\tt arXiv:1207.5795 [hep-ph]}}].
%%CITATION = ARXIV:1207.5795;%%

\bibitem{Romatschke:2005pm}
P.~Romatschke and R.~Venugopalan, {\it Collective {non-Abelian} instabilities
  in a melting color glass condensate},
  \href{http://dx.doi.org/10.1103/PhysRevLett.96.062302}{{\em Phys. Rev. Lett.}
  {\bf 96} (2006) 062302} [\href{http://arXiv.org/abs/hep-ph/0510121}{{\tt
  arXiv:hep-ph/0510121}}].
%%CITATION = HEP-PH/0510121;%%

\bibitem{Romatschke:2006nk}
P.~Romatschke and R.~Venugopalan, {\it The unstable {Glasma}},
  \href{http://dx.doi.org/10.1103/PhysRevD.74.045011}{{\em Phys. Rev.} {\bf
  D74} (2006) 045011} [\href{http://arXiv.org/abs/hep-ph/0605045}{{\tt
  arXiv:hep-ph/0605045}}].
%%CITATION = HEP-PH/0605045;%%

\bibitem{Lappi:2016ato}
T.~Lappi and J.~Peuron, {\it {Plasmon mass scale in classical nonequilibrium
  gauge theory}},  \href{http://dx.doi.org/10.1103/PhysRevD.95.014025}{{\em
  Phys. Rev.} {\bf D95} (2017)~no.~1 014025}
  [\href{http://arXiv.org/abs/1610.03711}{{\tt arXiv:1610.03711 [hep-ph]}}].
%%CITATION = ARXIV:1610.03711;%%

\bibitem{Krasnitz:2000gz}
A.~Krasnitz and R.~Venugopalan, {\it The initial gluon multiplicity in heavy
  ion collisions},  \href{http://dx.doi.org/10.1103/PhysRevLett.86.1717}{{\em
  Phys. Rev. Lett.} {\bf 86} (2001) 1717}
  [\href{http://arXiv.org/abs/hep-ph/0007108}{{\tt arXiv:hep-ph/0007108}}].
%%CITATION = HEP-PH/0007108;%%

\bibitem{Mace:2016svc}
M.~Mace, S.~Schlichting and R.~Venugopalan, {\it Off-equilibrium sphaleron
  transitions in the glasma},
  \href{http://dx.doi.org/10.1103/PhysRevD.93.074036}{{\em Phys. Rev.} {\bf
  D93} (2016) 074036} [\href{http://arXiv.org/abs/1601.07342}{{\tt
  arXiv:1601.07342 [hep-ph]}}].
%%CITATION = ARXIV:1601.07342;%%

\bibitem{Mueller:2016ven}
N.~Mueller, S.~Schlichting and S.~Sharma, {\it {Chiral magnetic effect and
  anomalous transport from real-time lattice simulations}},
  \href{http://dx.doi.org/10.1103/PhysRevLett.117.142301}{{\em Phys. Rev.
  Lett.} {\bf 117} (2016) 142301} [\href{http://arXiv.org/abs/1606.00342}{{\tt
  arXiv:1606.00342 [hep-ph]}}].
%%CITATION = ARXIV:1606.00342;%%

\bibitem{Kurkela:2012hp}
A.~Kurkela and G.~D. Moore, {\it {UV} cascade in classical {Yang-Mills}
  theory},  \href{http://dx.doi.org/10.1103/PhysRevD.86.056008}{{\em Phys.
  Rev.} {\bf D86} (2012) 056008} [\href{http://arXiv.org/abs/1207.1663}{{\tt
  arXiv:1207.1663 [hep-ph]}}].
%%CITATION = ARXIV:1207.1663;%%

\bibitem{Berges:2008zt}
J.~Berges, D.~Gelfand, S.~Scheffler and D.~Sexty, {\it Simulating plasma
  instabilities in {SU(3)} gauge theory},
  \href{http://dx.doi.org/10.1016/j.physletb.2009.05.008}{{\em Phys. Lett.}
  {\bf B677} (2009) 210} [\href{http://arXiv.org/abs/0812.3859}{{\tt
  arXiv:0812.3859 [hep-ph]}}].
%%CITATION = ARXIV:0812.3859;%%

\bibitem{Ipp:2010uy}
A.~Ipp, A.~Rebhan and M.~Strickland, {\it {Non-Abelian} plasma instabilities:
  {SU(3)} vs. {SU(2)}},
  \href{http://dx.doi.org/10.1103/PhysRevD.84.056003}{{\em Phys. Rev.} {\bf
  D84} (2011) 056003} [\href{http://arXiv.org/abs/1012.0298}{{\tt
  arXiv:1012.0298 [hep-ph]}}].
%%CITATION = ARXIV:1012.0298;%%

\bibitem{Berges:2017igc}
J.~Berges, M.~Mace and S.~Schlichting, {\it {Universal self-similar scaling of
  spatial Wilson loops out of equilibrium}},
  \href{http://dx.doi.org/10.1103/PhysRevLett.118.192005}{{\em Phys. Rev.
  Lett.} {\bf 118} (2017)~no.~19 192005}
  [\href{http://arXiv.org/abs/1703.00697}{{\tt arXiv:1703.00697 [hep-th]}}].
%%CITATION = ARXIV:1703.00697;%%

\bibitem{Davies:1987vs}
C.~T.~H. Davies, G.~G. Batrouni, G.~R. Katz, A.~S. Kronfeld, G.~P. Lepage,
  K.~G. Wilson, P.~Rossi and B.~Svetitsky, {\it {Fourier} acceleration in
  lattice gauge theories. 1. {Landau} gauge fixing},
  \href{http://dx.doi.org/10.1103/PhysRevD.37.1581}{{\em Phys. Rev.} {\bf D37}
  (1988) 1581}.
%%CITATION = PHRVA,D37,1581;%%

\bibitem{Li:2017iat}
M.~Li, {\it {Reexamining the gluon spectrum in the boost-invariant glasma from
  a semianalytic approach}},
  \href{http://dx.doi.org/10.1103/PhysRevC.96.064904}{{\em Phys. Rev.} {\bf
  C96} (2017)~no.~6 064904} [\href{http://arXiv.org/abs/1711.00409}{{\tt
  arXiv:1711.00409 [nucl-th]}}].
%%CITATION = ARXIV:1711.00409;%%

\bibitem{Gyulassy:2004zy}
M.~Gyulassy and L.~McLerran, {\it New forms of {QCD} matter discovered at
  {RHIC}},  \href{http://dx.doi.org/10.1016/j.nuclphysa.2004.10.034}{{\em Nucl.
  Phys.} {\bf A750} (2005) 30}
  [\href{http://arXiv.org/abs/nucl-th/0405013}{{\tt arXiv:nucl-th/0405013}}].
%%CITATION = NUCL-TH/0405013;%%

\bibitem{Berges:2007re}
J.~Berges, S.~Scheffler and D.~Sexty, {\it Bottom-up isotropization in
  classical-statistical lattice gauge theory},
  \href{http://dx.doi.org/10.1103/PhysRevD.77.034504}{{\em Phys. Rev.} {\bf
  D77} (2008) 034504} [\href{http://arXiv.org/abs/0712.3514}{{\tt
  arXiv:0712.3514 [hep-ph]}}].
%%CITATION = ARXIV:0712.3514;%%

\bibitem{Berges:2012ev}
J.~Berges, S.~Schlichting and D.~Sexty, {\it Over-populated gauge fields on the
  lattice},  \href{http://dx.doi.org/10.1103/PhysRevD.86.074006}{{\em Phys.
  Rev.} {\bf D86} (2012) 074006} [\href{http://arXiv.org/abs/1203.4646}{{\tt
  arXiv:1203.4646 [hep-ph]}}].
%%CITATION = ARXIV:1203.4646;%%

\bibitem{Berges:2013eia}
J.~Berges, K.~Boguslavski, S.~Schlichting and R.~Venugopalan, {\it Turbulent
  thermalization process in heavy-ion collisions at ultrarelativistic
  energies},  \href{http://dx.doi.org/10.1103/PhysRevD.89.074011}{{\em Phys.
  Rev.} {\bf D89} (2014) 074011} [\href{http://arXiv.org/abs/1303.5650}{{\tt
  arXiv:1303.5650 [hep-ph]}}].
%%CITATION = ARXIV:1303.5650;%%

\bibitem{Berges:2014bba}
J.~Berges, K.~Boguslavski, S.~Schlichting and R.~Venugopalan, {\it Universality
  far from equilibrium: From superfluid {Bose} gases to heavy-ion collisions},
  \href{http://dx.doi.org/10.1103/PhysRevLett.114.061601}{{\em Phys. Rev.
  Lett.} {\bf 114} (2015) 061601} [\href{http://arXiv.org/abs/1408.1670}{{\tt
  arXiv:1408.1670 [hep-ph]}}].
%%CITATION = ARXIV:1408.1670;%%

\bibitem{Lappi:2011ju}
T.~Lappi, {\it Gluon spectrum in the glasma from {JIMWLK} evolution},
  \href{http://dx.doi.org/10.1016/j.physletb.2011.08.011}{{\em Phys. Lett.}
  {\bf B703} (2011) 325} [\href{http://arXiv.org/abs/1105.5511}{{\tt
  arXiv:1105.5511 [hep-ph]}}].

\bibitem{Berges:2012cj}
J.~Berges and S.~Schlichting, {\it {The nonlinear glasma}},
  \href{http://dx.doi.org/10.1103/PhysRevD.87.014026}{{\em Phys. Rev.} {\bf
  D87} (2013) 014026} [\href{http://arXiv.org/abs/1209.0817}{{\tt
  arXiv:1209.0817 [hep-ph]}}].
%%CITATION = ARXIV:1209.0817;%%

\bibitem{Berges:2013fga}
J.~Berges, K.~Boguslavski, S.~Schlichting and R.~Venugopalan, {\it Universal
  attractor in a highly occupied {non Abelian} plasma},
  \href{http://dx.doi.org/10.1103/PhysRevD.89.114007}{{\em Phys. Rev.} {\bf
  D89} (2014) 114007} [\href{http://arXiv.org/abs/1311.3005}{{\tt
  arXiv:1311.3005 [hep-ph]}}].
%%CITATION = ARXIV:1311.3005;%%

\bibitem{Berges:2013lsa}
J.~Berges, K.~Boguslavski, S.~Schlichting and R.~Venugopalan, {\it Basin of
  attraction for turbulent thermalization and the range of validity of
  classical-statistical simulations},
  \href{http://dx.doi.org/10.1007/JHEP05(2014)054}{{\em JHEP} {\bf 05} (2014)
  054} [\href{http://arXiv.org/abs/1312.5216}{{\tt arXiv:1312.5216 [hep-ph]}}].
%%CITATION = ARXIV:1312.5216;%%

\bibitem{Berges:2015ixa}
J.~Berges, K.~Boguslavski, S.~Schlichting and R.~Venugopalan, {\it
  {Nonequilibrium fixed points in longitudinally expanding scalar theories:
  infrared cascade, Bose condensation and a challenge for kinetic theory}},
  \href{http://dx.doi.org/10.1103/PhysRevD.92.096006}{{\em Phys. Rev.} {\bf
  D92} (2015)~no.~9 096006} [\href{http://arXiv.org/abs/1508.03073}{{\tt
  arXiv:1508.03073 [hep-ph]}}].
%%CITATION = ARXIV:1508.03073;%%

\bibitem{Moore:1996qs}
G.~D. Moore, {\it Motion of {Chern-Simons} number at high temperatures under a
  chemical potential},
  \href{http://dx.doi.org/10.1016/S0550-3213(96)00445-2}{{\em Nucl. Phys.} {\bf
  B480} (1996) 657} [\href{http://arXiv.org/abs/hep-ph/9603384}{{\tt
  arXiv:hep-ph/9603384 [hep-ph]}}].
%%CITATION = HEP-PH/9603384;%%

\bibitem{Kurkela:2016mhu}
A.~Kurkela, T.~Lappi and J.~Peuron, {\it {Time evolution of linearized gauge
  field fluctuations on a real-time lattice}},
  \href{http://dx.doi.org/10.1140/epjc/s10052-016-4523-9}{{\em Eur. Phys. J.}
  {\bf C76} (2016)~no.~12 688} [\href{http://arXiv.org/abs/1610.01355}{{\tt
  arXiv:1610.01355 [hep-lat]}}].
%%CITATION = ARXIV:1610.01355;%%

\end{thebibliography}\endgroup
% \bibliography{master}
\bibliographystyle{JHEP-2modlong}

\end{document}